\documentclass[prb,floatfix,twocolumn,showpacs,amsmath,amssymb]{revtex4}
\usepackage{epstopdf}
\usepackage{epsfig}
\usepackage{dcolumn}
\usepackage{bm}
\usepackage{color}

\usepackage[normalem]{ulem}

\usepackage{chngcntr}

\allowdisplaybreaks

\begin{document}

\title{Emergent lattices with geometrical frustration in doped extended Hubbard models}
\author{%
Ryui Kaneko,$^1$ 
Luca F. Tocchio,$^{2}$
Roser Valent\'i,$^{1}$ 
and Claudius Gros$^{1}$}
\affiliation{
$^{1}$Institute for Theoretical Physics, University of Frankfurt,
       Max-von-Laue-Stra{\ss}e 1, D-60438 Frankfurt a.M., Germany \\
$^{2}$Democritos National Simulation Center, Istituto Officina dei
       Materiali del CNR, and SISSA-International School for Advanced
       Studies, Via Bonomea 265, I-34136 Trieste, Italy
            }

\date{\today}

\begin{abstract}
Spontaneous charge ordering occurring in correlated systems may be 
considered as a possible route to generate effective lattice 
structures with unconventional couplings. For this purpose
we investigate the phase diagram of doped extended Hubbard models 
on two lattices: (i) the honeycomb lattice with on-site $U$
and nearest-neighbor $V$ Coulomb interactions at $3/4$ filling 
($n=3/2$) and (ii) the triangular lattice with on-site $U$, 
nearest-neighbor $V$, and next-nearest-neighbor $V'$ Coulomb 
interactions at $3/8$ filling ($n=3/4$). We consider various 
approaches including mean-field approximations, perturbation theory,
and variational Monte Carlo.
For the  honeycomb case (i), charge order induces an effective 
triangular lattice at large values of $U/t$ and $V/t$,
where $t$ is the nearest-neighbor hopping integral. The 
nearest-neighbor spin exchange interactions on this effective 
triangular lattice are antiferromagnetic in most of the phase diagram, 
while they become ferromagnetic when $U$ is much larger than $V$.
At $U/t\sim (V/t)^3$,  ferromagnetic and antiferromagnetic 
exchange interactions nearly cancel out, leading to a system with
four-spin ring-exchange interactions.
On the other hand, for the triangular case (ii) 
at large $U$ and  finite $V'$,
we find no charge order for small $V$, an effective kagome lattice
 for intermediate $V$, and one-dimensional charge order for large  $V$.
 These results indicate that Coulomb interactions  induce [case (i)]  or enhance [case(ii)]
emergent geometrical frustration of the spin degrees of freedom 
in the system, by forming charge order.
\end{abstract}


\maketitle


\section{Introduction}

Correlated systems with competing on-site and intersite
Coulomb interactions~\cite{imada1998} and fillings away from 
one electron per site ($n=1$, half filling)
are presently a subject of intensive investigation due to the
appearance  of complex phases such as unconventional charge and 
magnetic orders. These systems become even more challenging when novel
lattice structures emerge out of the original lattice in the region 
of strong correlations~\cite{baskaran2016}.
This phenomenon is often found in geometrically 
frustrated systems, such as triangular and kagome lattices. 

On the 
triangular lattice, for instance, large on-site $U$ and nearest-neighbor 
$V$ Coulomb interactions generate effective honeycomb and enlarged 
triangular lattices at $1/3$ filling ($n=2/3$) by inducing charge 
disproportionation~\cite{watanabe2005,tocchio2014,kaneko2016}.
When $V\gtrsim U/3$  the system tends to create a honeycomb lattice 
of empty sites and an enlarged triangular lattice of doubly occupied 
sites, while at smaller ratios of $V/U$ the system evolves into a 
honeycomb lattice of singly occupied sites 
with long-range 
antiferromagnetic order.
A similar charge ordered state with noncollinear magnetic order 
has also been
proposed in the Kondo lattice system~\cite{reja2015}.
While these states are insulating,
such exotic charge and magnetic orders become metallic away from 
the commensurate filling~\cite{tocchio2014}. Furthermore, at quarter 
filling ($n=1/2$), metallic states, named pinball liquids, have been 
also recently proposed~\cite{hotta2006,miyazaki2009,canocortes2011,merino2013}. 
They are characterized by a three-sublattice structure, in which the 
carriers of one sublattice are essentially localized (pins), with the 
remaining charges (balls) building an itinerant liquid on the interstitials.
Recently,  other mechanisms than direct
charge disproportionation have been also proposed to generate
new lattice structures such as the emergence
of a kagome lattice via spontaneous ferrimagnetic
order coexisting
with a $\sqrt{3}\times\sqrt{3}$ charge order
pattern in a triangular Kondo lattice~\cite{akagi2015}.

Similarly, on the kagome lattice, large values of $U$ and $V$ have been 
discussed to induce nearly isolated six-site rings and an enlarged kagome 
lattice at $1/3$ filling ($n=2/3$)~\cite{wen2010,ferhat2014,pollmann2014}.
Specifically, when $U,V>0$ and $t=0$, each corner-sharing triangle 
possesses
charge order characterized by two singly occupied
sites and an empty site.
The 
empty site
randomly sits on one 
of the three vertices of a triangle, which gives macroscopic charge degeneracy.
Nonzero hopping $t$ lifts the charge degeneracy and appears to stabilize a
$\sqrt{3}\times\sqrt{3}$ charge pattern, whose unit cell contains nine
sites~\cite{wen2010,ferhat2014,pollmann2014}.
Recently, by mapping the system into a hard-core boson Hamiltonian,
a topological liquid  was also proposed~\cite{roychowdhury2015}.

Reported realizations of such emergent lattices are for instance the 
generation of a honeycomb
structure through charge disproportion  in the
metallic magnet ${\rm AgNiO_2}$~\cite{wawrzynska2007,wawrzynska2008}
or the appearance of
 effective spin-$1/2$ chains 
in the heavy-fermion spinel ${\rm LiV_2O_4}$~\cite{fulde2001,fulde2002}.

Actually, even when the lattice structures themselves are not geometrically 
frustrated, such a formation of new lattices is possible due to  strong 
electron correlations. For example, in the cubic lattice, a staggered 
$(\pi,\pi,\pi)$ charge order generates doubled face-centered-cubic 
lattices~\cite{hayami2014}. Moreover, effective spin and charge interactions 
in such systems may acquire additional geometrical frustrations. Indeed, 
unconventional noncoplanar magnetic orders have been proposed in the 
periodic Anderson model on the cubic lattice at $n=3/2$ filling~\cite{hayami2014}.

One of the most discussed bipartite lattices in two-dimensional systems is
the honeycomb lattice. While correlation effects on the honeycomb lattice
have been intensively discussed in the past, many studies focused mostly at 
half filling~\cite{meng2010,sorella2012,raghu2008,weeks2010,capponi2015,motruk2015,scherer2015,kurita2015}.  
Castro {\it et al.}~\cite{castro2011} and 
Grushin {\it et al.}~\cite{grushin2013} studied extensively 
the phase diagram in honeycomb systems for arbitrary filling; however, 
they considered only  the spinless Hubbard model.

In this work, we investigate the possible emergence of correlation-induced 
new lattice structures both in a bipartite and in a geometrically frustrated 
lattice at fillings that, to our knowledge, have not been investigated before.
In particular, we perform an extensive analysis of the spinful extended Hubbard 
model on honeycomb and triangular lattices away from half filling, and focus 
on the interplay between on-site $U$ and intersite nearest-
and next-nearest Coulomb interactions $V$ and $V'$, respectively. By using 
the Hartree-Fock approximation, as well as perturbation theory and the 
variational Monte Carlo method, we find that a triangular structure emerges 
from charge order on the honeycomb lattice for large values of $U$ and $V$ 
at $3/4$ filling ($n=3/2$, three electrons per two sites on average). Charge-poor sites possess spin degrees of freedom, 
and their spin correlations are found to be antiferromagnetic in most of the 
phase diagram, while they become ferromagnetic when $U$ is much larger than $V$.

On the other hand, for the triangular lattice with $U$, $V$, and $V'$
interactions at $3/8$ filling ($n=3/4$, three electrons per four
sites on average), considering large values of $U$ and 
a finite $V'$ (we set $V' = V/5$), we find that the system shows rich charge orders:
a kagome structure emerges for intermediate values of $V$,
while a one-dimensional structure is stabilized for large values of $V$.
Both examples show an enhancement of  geometrical frustration, 
from the honeycomb to the triangular lattice in the former case, and 
from the triangular to the kagome in the latter one.

The paper is organized as follows.
In Sec.~\ref{sec:honeycomb}, we present the extended Hubbard model 
on the honeycomb lattice and show the possible phases of the model 
as a function of $U$ and $V$ obtained with the
 Hartree-Fock approximation and with variational Monte Carlo. We also discuss how $U$ and $V$ determine 
magnetic patterns of the emergent charge ordered states by means of  
perturbation theory.  In Sec.~\ref{sec:triangular}, we present
variational Monte Carlo results for the 
extended Hubbard model on the triangular lattice and discuss possible 
phases of the model as a function of $V$ for large $U$ and $V'= V/5$.
Finally, in Sec.~\ref{sec:conclusions}, we draw our conclusions.

\section{Emergent triangular structure on a honeycomb system}
\label{sec:honeycomb}

\begin{figure}[t]
\includegraphics[width=0.9\columnwidth]{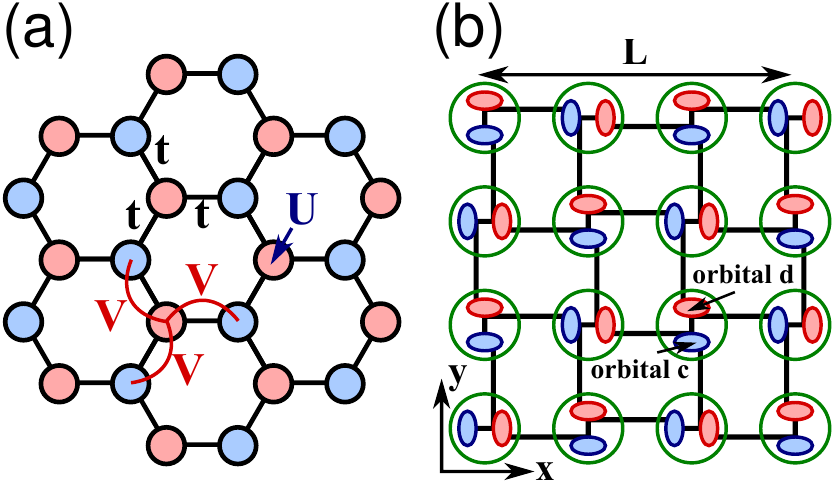}
\caption{(Color online) The extended Hubbard model with 
hopping $t$, on-site Coulomb interaction $U$, and nearest-neighbor
Coulomb interaction $V$, Eq.~(\ref{eq:hubbard}) and Eq.~(\ref{eq:hubbard_2orbital}),
on (a) the honeycomb lattice  and (b) its equivalent two-band 
representation, Eq.~(\ref{eq:hubbard_2orbital}). Blue and red circles denote
orbitals $c$ and $d$, respectively.}
\label{fig:lattice}
\end{figure}

\subsection{Extended Hubbard model on a honeycomb lattice}

We consider an extended Hubbard model on the isotropic honeycomb lattice
[see Fig.~\ref{fig:lattice}(a)] where the Hamiltonian is given as
\begin{eqnarray}
\label{eq:hubbard}
 H &=&
  - t \sum_{\langle i,j\rangle,\sigma}
    c^{\dagger}_{i,\sigma} c^{\phantom{\dagger}}_{j,\sigma}
  + \mathrm{h.c.}
\nonumber
\\
 &&
  + U \sum_{i}
    n_{i,\uparrow} n_{i,\downarrow}
  + V \sum_{\langle i,j\rangle}
    n_{i} n_{j};
\end{eqnarray}
$t$ denotes the hopping parameter, and
$U$ and $V$ are the on-site and nearest-neighbor Coulomb interaction,
respectively.
Hereafter, we investigate repulsive Coulomb interactions ($U,V\ge 0$),
and focus on $3/4$ filling ($n=3/2$).
We note that on the honeycomb lattice
 $3/4$  and $1/4$ fillings are equivalent 
via the particle-hole transformation.

\begin{figure*}[t]
\includegraphics[height=0.65\columnwidth]{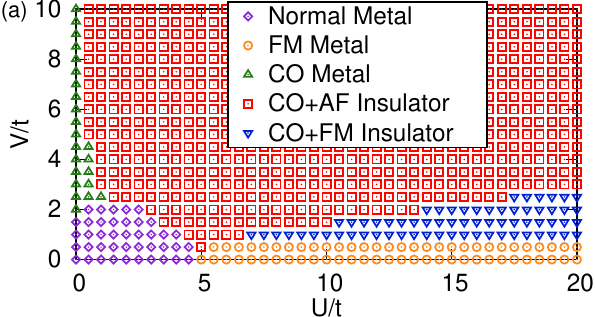}\hspace{10ex}
\includegraphics[height=0.65\columnwidth]{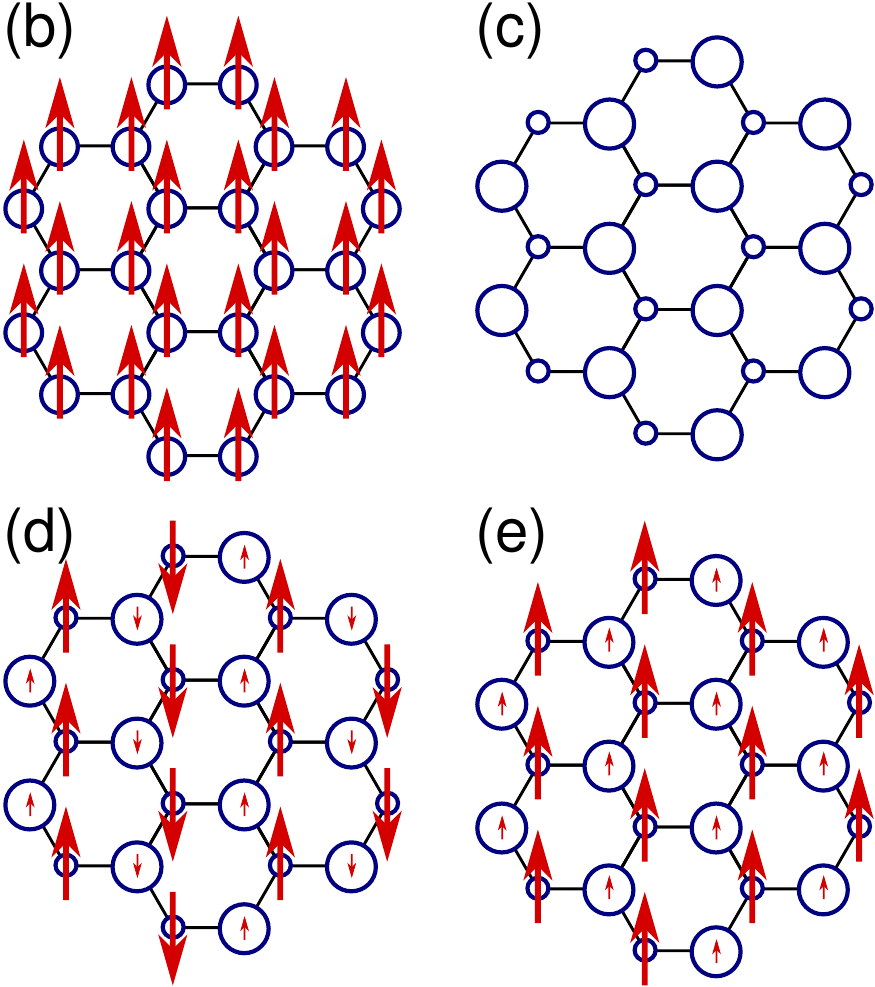}
\caption{(Color online)
(a) Hartree-Fock phase diagram of the Hubbard model at $n=3/2$, 
see Eq.~(\ref{eq:hubbard}), on the honeycomb lattice.
(b) Illustration of the ferromagnetic metallic state 
(FM metal), with one up-spin and half a down-spin (on average) per
site.
(c) Charge ordered metallic state (CO metal), with alternating 
doubly and singly occupied site (large/small circles).
(d) Charge ordered antiferromagnetic insulating state (CO+AF
insulator).
(e) Charge ordered ferromagnetic insulating state (CO+FM insulator).}
\label{fig:MF_phase_diag}
\end{figure*}

This model, being defined on a lattice with two sites per
unit cell, can be also regarded as a two-band (two-orbital)
Hubbard model whose hoppings
connect only different orbitals [see Fig.~\ref{fig:lattice}(b)],
\begin{eqnarray}
\label{eq:hubbard_2orbital}
 H &=&
  - t \sum_{i,\sigma}
  \left(
    d^{\dagger}_{i,\sigma} c^{\phantom{\dagger}}_{i,\sigma}
  + d^{\dagger}_{i,\sigma} c^{\phantom{\dagger}}_{i+\bm{e}_x,\sigma}
  + d^{\dagger}_{i,\sigma} c^{\phantom{\dagger}}_{i+\bm{e}_y,\sigma}
  + \mathrm{h.c.}
  \right)
\nonumber
\\
 &&
  + U \sum_{i}
  \left(
    n_{i,\uparrow}^{c} n_{i,\downarrow}^{c}
  + n_{i,\uparrow}^{d} n_{i,\downarrow}^{d}
  \right)
\nonumber
\\
 &&
  + V \sum_{i}
  \left(
    n_{i}^{d} n_{i}^{c}
  + n_{i}^{d} n_{i+\bm{e}_x}^{c}
  + n_{i}^{d} n_{i+\bm{e}_y}^{c}
  \right),
\end{eqnarray}
Both representations of the Hamiltonian are equivalent and
we will make use of the latter representation for computational
purposes.

\subsection{Mean-field phase diagram}

In order to investigate the interplay between charge and magnetic 
order, we start with a mean-field analysis of the above presented
Hamiltonian.  Figure~\ref{fig:MF_phase_diag} (a) shows the  
$U$-$V$ phase diagram of the honeycomb model of
Eq.~(\ref{eq:hubbard_2orbital})  at $3/4$ filling 
obtained with the restricted Hartree-Fock method (as
explained in Appendix~\ref{sec:Hartree-Fock}).
For simplicity, we have restricted ourselves to coplanar
magnetic order patterns.

\begin{figure}[t]
\includegraphics[width=0.4\columnwidth]{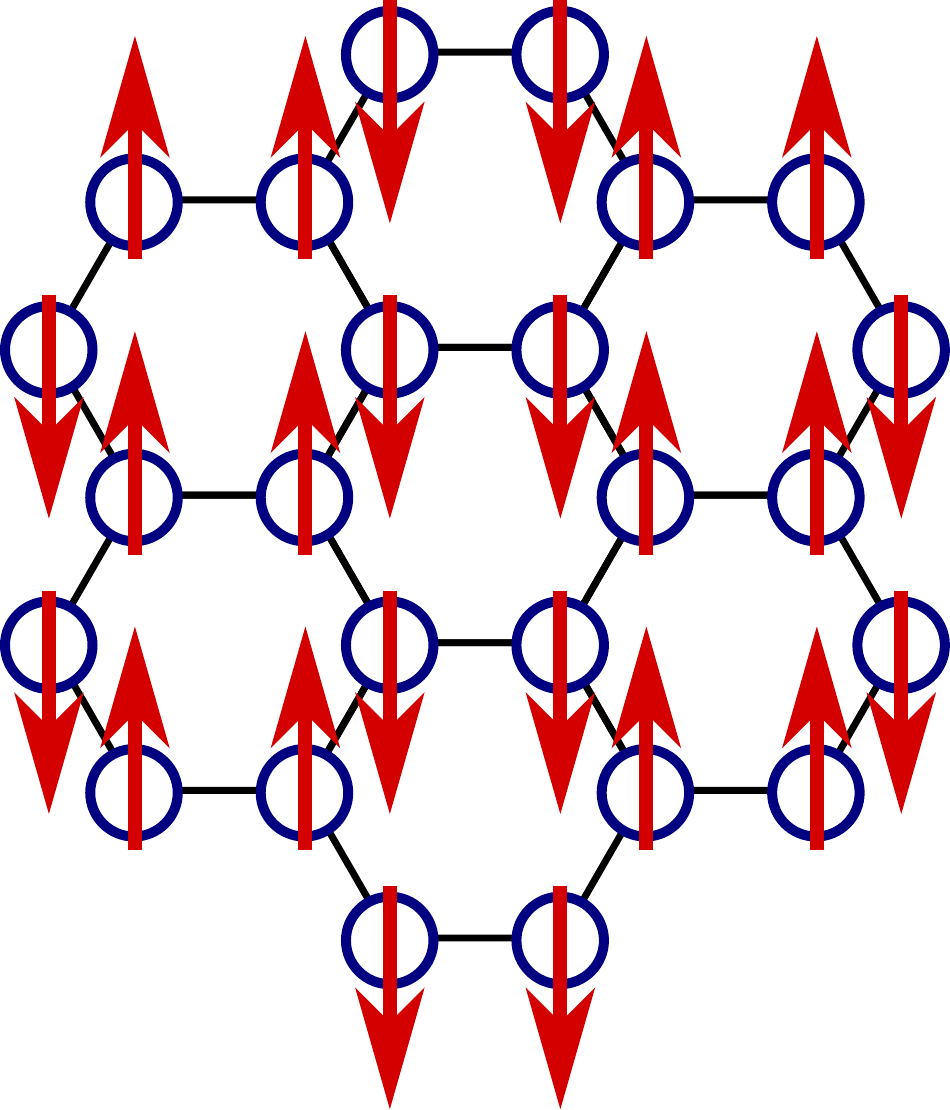}
\caption{(Color online)
Metastable state found at $V=0$.
Antiferromagnetic (up-up-down-down) insulator without charge order.
}
\label{fig:V0_states}
\end{figure}

\begin{figure}[t]
\includegraphics[width=0.9\columnwidth]{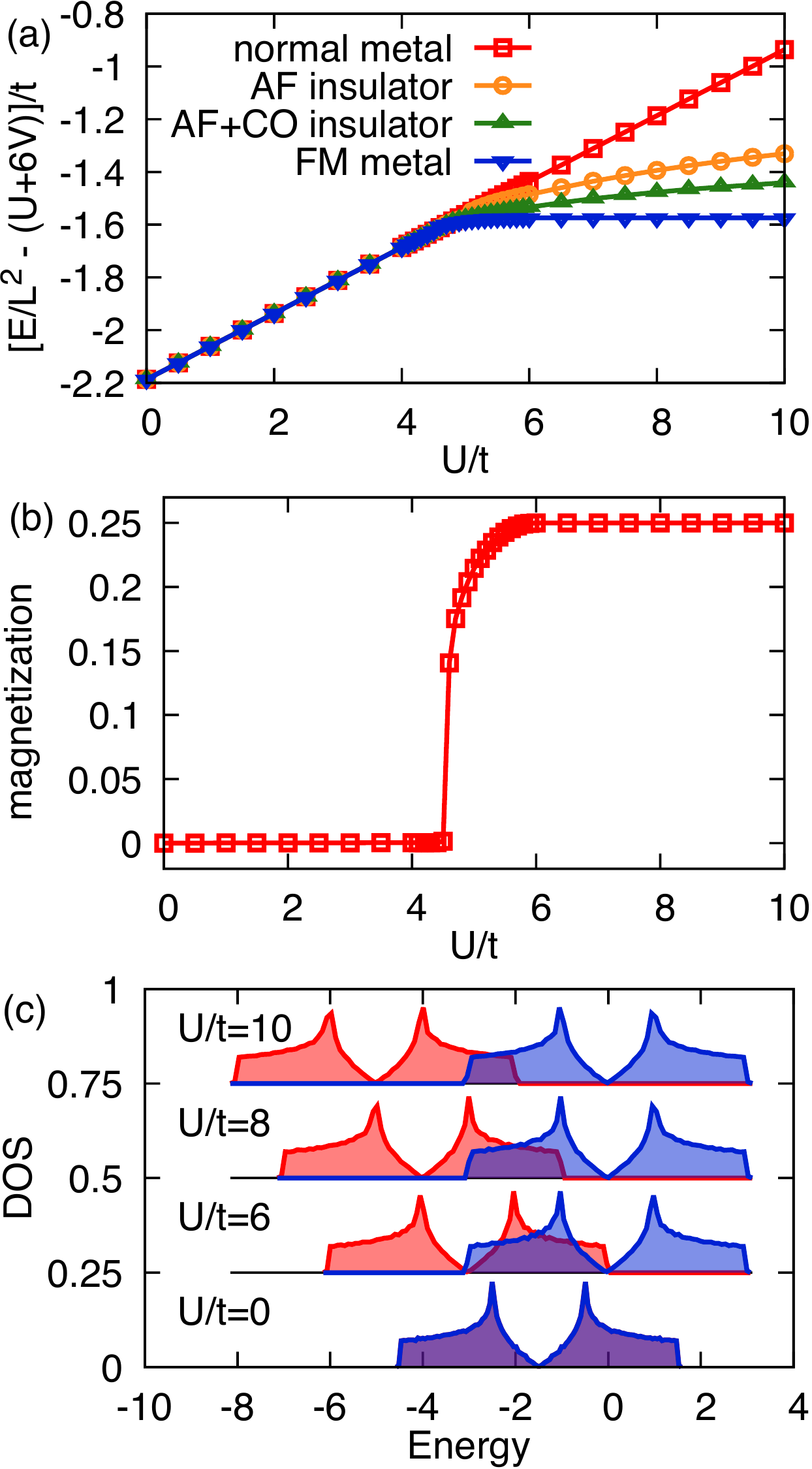}%
\caption{(Color online)
Results for the honeycomb lattice at $3/4$ filling for
$V=0$ obtained by the Hartree-Fock method.
(a) Energy of each state as a function of $U/t$.
(b) Magnetization of the metallic ground state (normal and FM).
(c) Density of states for the ground-state FM metal at each value of $U/t$.
The Fermi level is set to $0$. When the up-spin band is completely
below the Fermi level, the state becomes semimetallic.}
\label{fig:V0_MF}
\end{figure}

\begin{figure}[t]
\includegraphics[width=0.9\columnwidth]{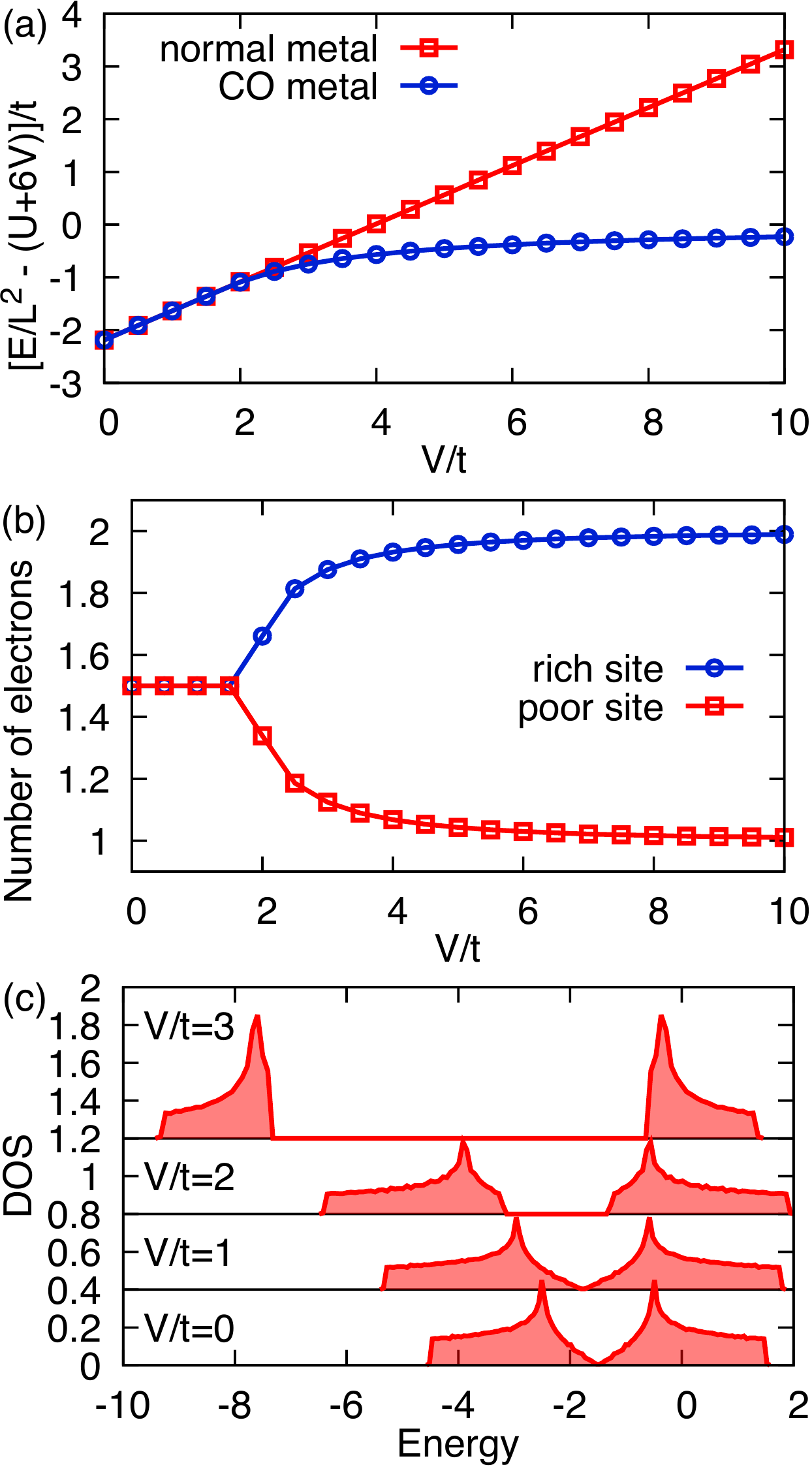}%
\caption{(Color online)
Results for the honeycomb lattice at $3/4$ filling
for  $U=0$ obtained by the Hartree-Fock
method.
(a) Energy of metallic states as a function of $V/t$.
(b) The number of electrons per each sublattice for the metallic ground states
(with and without charge order).
(c) Density of states for the ground-state metallic states.
The Fermi level is set to $0$.}
\label{fig:U0_MF}
\end{figure}

In the absence of nearest-neighbor Coulomb interaction $V$
(along $V=0$)
we find four ground-state candidates:
normal metal, ferromagnetic metal [Fig.~\ref{fig:MF_phase_diag}(b)],
 and antiferromagnetic insulator with
and without charge order 
[see Fig.~\ref{fig:MF_phase_diag}(d) and Fig.~\ref{fig:V0_states}].
As shown in Fig.~\ref{fig:V0_MF}(a), the energies of 
 antiferromagnetic and charge ordered states are
always higher than those of normal and ferromagnetic metal states.
A continuous phase transition from normal to ferromagnetic metal occurs at $U/t\sim 5$,
as shown in Fig.~\ref{fig:V0_MF}(b).
This ferromagnetic metal [Fig.~\ref{fig:MF_phase_diag} (b)]
is consistent with the result obtained by
Hanisch {\it et al}~\cite{hanisch1997}.
When $U/t\gtrsim 6$, spins are fully polarized.
In the ferromagnetic state at $3/4$ filling,
the  up-spin lower band is fully occupied while
the down-spin upper band is half filled and the
density of states (DOS) is
 zero at the Fermi energy indicating
a semimetallic behavior. Figure~\ref{fig:V0_MF}(c)
shows  the DOS as a function of $U$.

\begin{figure}[t]
\includegraphics[width=0.9\columnwidth]{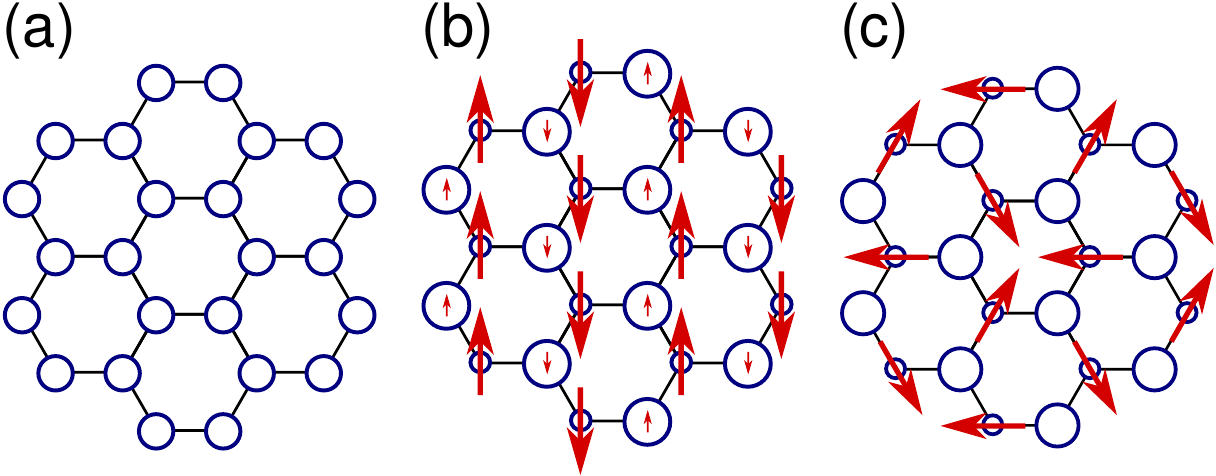}%
\caption{(Color online)
States found 
for $U=V$.
(a) Normal metal without charge and magnetic order.
(b) Charge ordered stripe antiferromagnetic insulator.
(c) Charge ordered 120$^\circ$ antiferromagnetic insulator.
The energy of the stripe antiferromagnetic state is found to be the lowest one.
}
\label{fig:UV_states}
\end{figure}

\begin{figure}[t]
\includegraphics[width=0.8\columnwidth]{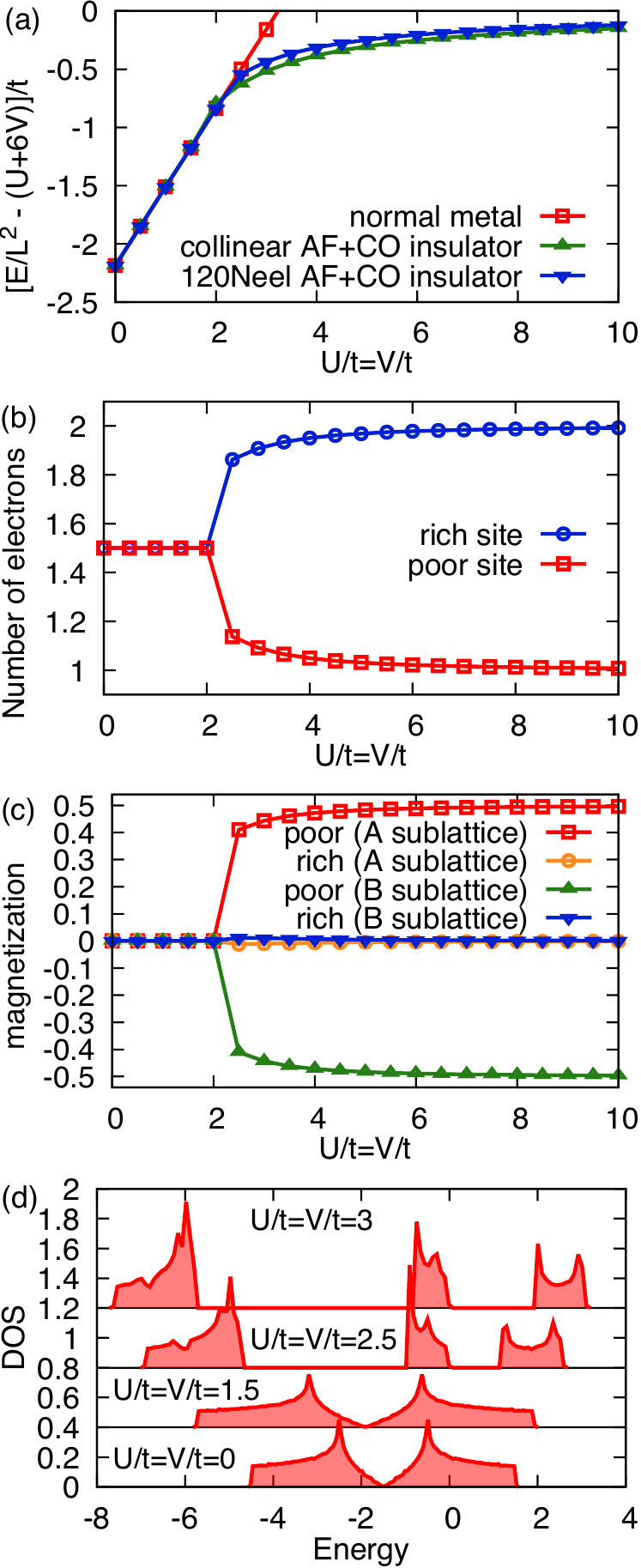}%
\caption{(Color online)
Results for the honeycomb lattice at $3/4$ filling
for $U=V$ obtained by the Hartree-Fock
method.
(a) Energy of each state as a function of $U=V$.
(b) The number of electrons per each sublattice for ground states
(normal metal and charge ordered collinear AF state).
(c) Magnetization of ground states.
Charge-poor (charge-rich) sites show large (small) magnetization.
(d) Density of states for the ground state.
The Fermi level is set to $0$.}
\label{fig:UV_MF}
\end{figure}

On the other hand, in the absence of on-site Coulomb interaction $U$
(along $U=0$) and for finite $V$,
within the $2\times 2$
sublattice structure, we obtain a staggered charge order state,
where $c$-orbital sites are charge-rich
($n_c=n_{c\uparrow}+n_{c\downarrow}>3/2$)
while $d$-orbital sites are charge-poor
($n_d=n_{d\uparrow}+n_{d\downarrow}<3/2$),
as shown in Fig.~\ref{fig:MF_phase_diag}(c).
In the absence of on-site Coulomb interaction ($U=0$),
this charge ordered state does not have any magnetic order.
As shown in Fig.~\ref{fig:U0_MF}, we find a phase transition
from the nonmagnetic metal to the charge ordered metal
at $V/t\sim 2$
which is stabilized by splitting the upper and lower bands.
This state is metallic since the upper band is
always half filled for $n=3/2$.

We  consider now the case of large $U$ and $V$ values, where
charge order is expected to generate complex magnetic orders.
As shown in Fig.~\ref{fig:MF_phase_diag} (a), when both $U$ and $V$ are
large, we find a charge ordered antiferromagnetic insulator. 
It has a rich-poor-rich-poor type charge pattern, and
the charge-poor sites form an emergent triangular structure. 
Magnetic order is found to be collinear
and shows stripe order  [Fig.~\ref{fig:MF_phase_diag} (d)].
On the other hand, when $U$ is much larger than $V$, 
a charge ordered ferromagnetic insulating phase 
[Fig.~\ref{fig:MF_phase_diag} (e)] appears.
It also has triangular-like charge order,
and charge-poor sites show dominant ferromagnetic order.

In order to investigate the possible antiferromagnetic 
patterns on the emergent triangular structure, we consider 
the collinear and the commensurate spiral state with $120^\circ$ order 
 of Fig.~\ref{fig:UV_states} along the $U=V$ line of the phase diagram.
We note that, in general, magnetic states may show incommensurate
coplanar spiral order or noncoplanar order in doped Hubbard
models~\cite{pasrija2016}. However, here we restrict ourselves to the  coplanar case.
As shown in Fig.~\ref{fig:UV_MF}, the stripe antiferromagnetic 
charge ordered state is found to have a
lower energy than the 120$^\circ$ N\'eel ordered state,
although the energy difference becomes extremely small as $U$ and $V$
increase.

The stability of the collinear phase should be induced
by second-order processes where a doubly occupied site is formed 
in the charge-poor sublattice, after the hopping of one electron 
from the charge-rich sublattice. Indeed, the hopping of one electron 
from a doubly occupied site to a singly occupied 
neighboring one is favored when collinearity holds, even for large values of $U$ and $V$. 
Since the intermediate state costs an energy $2V$, 
see for instance the first part of the process in Fig.~\ref{fig:UV_Jeff_Jnn},
the energy of the second-order process scales as $t^2/V$, 
as confirmed by the Hartree-Fock calculations for both stripe 
and 120$^\circ$ N\'eel antiferromagnetic charge ordered states
in the insulating phase. Moreover, as discussed in Appendix~\ref{sec:Perturbation}, 
effective next-neighbor exchange couplings on the emergent triangular lattice 
are generated for moderately large values of $U$ and $V$, favoring 
collinear orderings~\cite{jolicoeur1990,chubukov1992}.
All these contributions may then break down the 120 N\'{e}el spin state,
and induce the observed collinear pattern.

\subsection{Spin correlation in charge ordered states}
\label{sec:spin_correlations}

\begin{figure}[t]
\includegraphics[width=0.9\columnwidth]{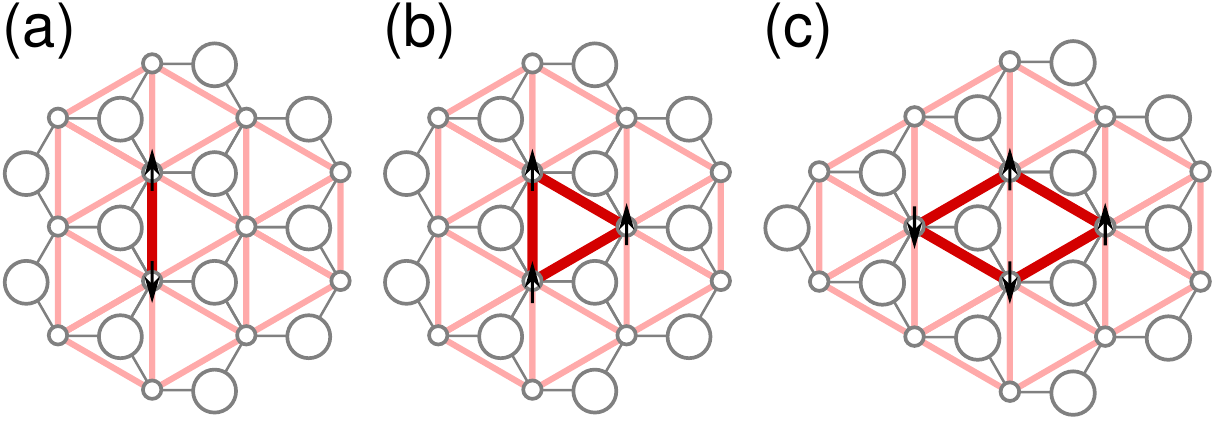}
\caption{(Color online)
Illustration of effective (a) two-, (b) three- spin interaction terms of 
Eqs.~(\ref{eq:H_spin}) and (\ref{eq:H_perm}) and (c) 
a four-site spin ring exchange.
}
\label{fig:UV_2_3_4_interaction}
\end{figure}

In the previous section, we showed that when correlations 
generate staggered charge order patterns on the honeycomb 
lattice, charge-rich and charge-poor sites form triangular 
lattices, respectively.
At $3/4$ filling, charge-rich sites contain two electrons on
average, while charge-poor sites contain one electron on average
with spin degrees of freedom.
In order to investigate how magnetic order appears in this limit,
we apply  perturbation theory to obtain an effective spin
Hamiltonian.

At the lowest order, the effective low-energy spin Hamiltonian on the
 triangular lattice (see Fig.~\ref{fig:UV_2_3_4_interaction}) contains
a spin exchange interaction and a three-particle permutation term, namely,
\begin{equation}
H_{\rm spin} = J_1\sum_{\langle i,j \rangle_1} \bm{S}_i\cdot \bm{S}_{j}
\label{eq:H_spin}
\end{equation}
and
\begin{equation}
 H_{\rm perm} = K_3\sum_{\triangle} (P_3 + P_3^{-1}).
\label{eq:H_perm}
\end{equation}
(Further details are given in Appendix~\ref{sec:Perturbation}.)
Here, the sum is taken over all nearest-neighbor sites
denoted by $\langle i,j \rangle_1$
for $H_{\rm spin}$,
while it is taken over  all triangles, which connect charge-poor sites, 
for $H_{\rm perm}$, as illustrated in Fig.~\ref{fig:UV_2_3_4_interaction}.
The symbol $P_n$ denotes a cyclic permutation operator.

\begin{figure*}[t]
\includegraphics[width=0.95\textwidth]{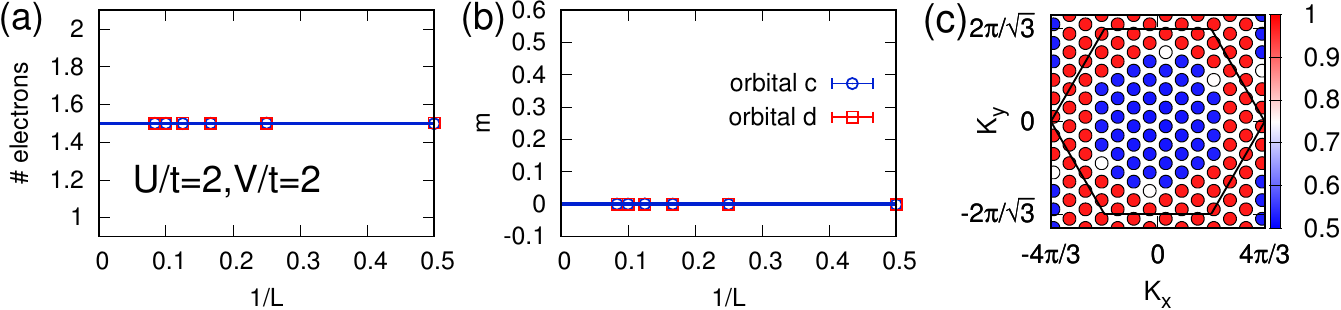}\newline
\includegraphics[width=0.95\textwidth]{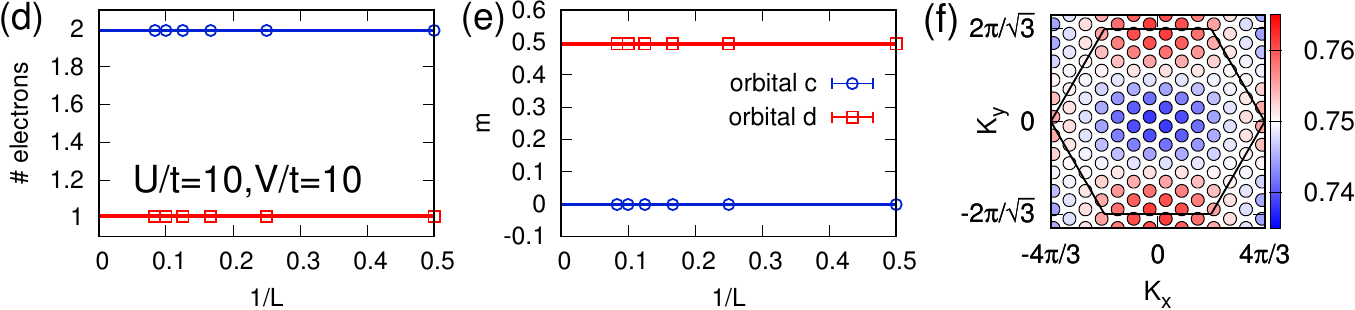}
\caption{(Color online)
Variational Monte Carlo (symbols) and Hartree-Fock (lines) 
results for the nonmagnetic metallic state ($U/t=V/t=2$, 
top row) and the charge-ordered antiferromagnetic
insulating state [$U/t=V/t=10$, bottom row; as illustrated
in Fig.~\ref{fig:MF_phase_diag}(d)]. In the first column
[panels (a) and (d)] the number of electrons for the two
sites making up the unit cell are given [denoted as $c$ and
$d$ orbitals; compare Eq.~(\ref{eq:hubbard_2orbital})]. In 
the second column
[panels (b) and (e)] the respective sublattice magnetizations.
In the last column [panels (c) and (f)] the respective momentum 
distributions $n(k)$, as evaluated for $200$ sites (using VMC) are presented.
The hexagon denotes the Brillouin zone of the honeycomb lattice.}
\label{fig:UV_VMC}
\end{figure*}

By considering  virtual hopping processes,
as shown in Fig.~\ref{fig:UV_Jeff_Jnn},
the exchange interaction $J_1$ can be evaluated as a function of $t$, $U$, and $V$,
namely, $J_1=c_1t^4/(V^2U)$ with a positive constant $c_1=1$.
This results in antiferromagnetic spin correlations.

Similarly, the coefficient $K_3$ in the permutation terms can be evaluated
by considering six cyclic processes for right-pointing triangles 
that lie inside the hexagons ($K_3^{\triangleright}$) and for left-pointing 
triangles that connect three hexagons ($K_3^{\triangleleft}$). 
In Appendix~\ref{sec:Perturbation}
 we show in Fig.~\ref{fig:UV_Jeff_Jr3}
one of the virtual processes generating $K_3^{\triangleright}$, which
does not require the formation of intermediate double-occupied sites. 
This coefficient survives even for $U=\infty$, namely, $K_3^{\triangleright}=-d_3t^6/V^5$
with a positive constant $d_3$.
When $U<\infty$, the formation of intermediate double-occupied sites leads to
other six cyclic process  in $K_3^{\triangleleft}$ and in
$K_3^{\triangleright}$. 
Since $P_3$ can be mapped to two-spin exchange operators~\cite{roger1983},
these permutation terms finally result in ferromagnetic exchange interactions.
The effective spin Hamiltonian is given by
\begin{equation}
H = J_1^{\rm eff} \sum_{\langle i,j \rangle_1} 
\bm{S}_i\cdot \bm{S}_{j}
\end{equation}
with $J_1^{\rm eff} = J_1+2K_3^{\triangleright}+2K_3^{\triangleleft}$.

In the limit $(V/t)^3\gg U/t$, antiferromagnetic spin 
correlations become relevant ($J_1\gg K_3$), and hence 
$J_1^{\rm eff}$ becomes antiferromagnetic. On the other 
hand, when $U/t\gg (V/t)^3$, antiferromagnetic spin 
correlations are suppressed
($K_3\gg J_1$), leading to a ferromagnetic $J_1^{\rm eff}$. This
is consistent with the results in the previous section.

Furthermore, when $(V/t)^3\sim U/t$, ferromagnetic $K_3$ and 
antiferromagnetic $J_1$ nearly cancel out. In this case, higher 
order processes in the perturbation theory become relevant. 
One of the dominant terms is a four-spin ring-exchange interaction
$K_4 (P_4 + P_4^{-1})$ on the effective triangular lattice,
as shown in Figs.~\ref{fig:UV_2_3_4_interaction} (c) and \ref{fig:UV_Jeff_Jr4}.
This may induce exotic spin liquid
states~\cite{motrunich2005,grover2010,holt2014}
or chiral magnetic order~\cite{korshunov1993,kubo1997}.
Moreover, the effective energy scale is extremely small ($|J_1^{\rm
eff}|\sim |K_4|\sim t^8/V^7$), which induces highly degenerate low-energy
states.

\begin{figure}[t]
\includegraphics[width=0.9\columnwidth]{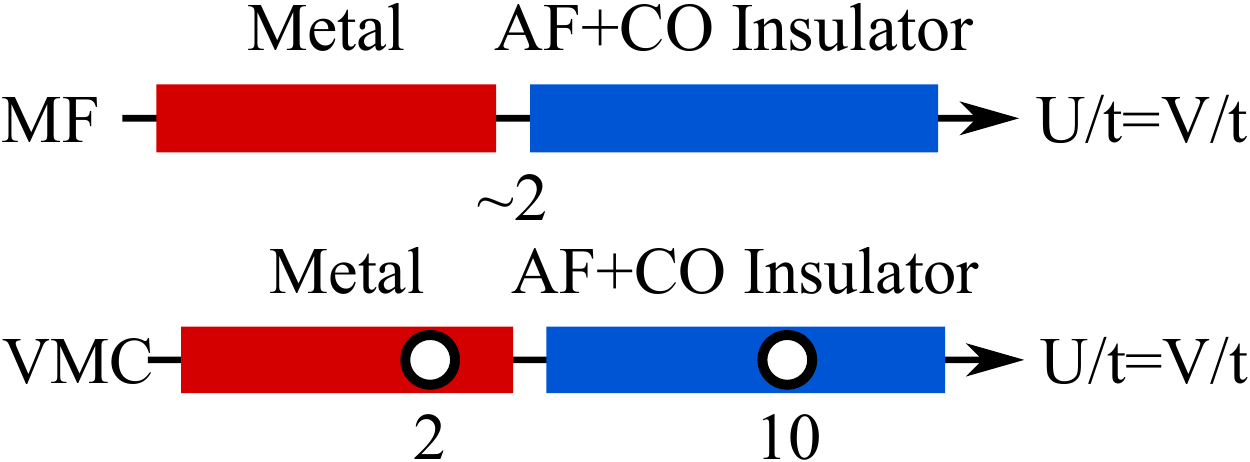}%
\caption{(Color online) Schematic phase diagram for the
honeycomb lattice at $3/4$ filling for $U=V$ obtained 
using Hartree-Fock (MF, top) and variational Monte
Carlo (VMC, bottom).}
\label{fig:UV_PD}
\end{figure}

\subsection{Charge ordering vs.\ phase separation}

The presence of charge order is a necessary condition
for the validity of the perturbation expansion discussed
in Sec.~\ref{sec:spin_correlations}. The mean-field approach tends
however to overestimate the stability range of ordered phases
and to underestimate the stability of nonordered metallic
phases stabilized in turn by quantum fluctuations. In what 
follows, we investigate the stability of the ordered state 
performing variational Monte Carlo simulations. We prepare the initial 
states by choosing the charge ordered states found in
the restricted Hartree-Fock method and then optimize the variational parameters. 
The details of the method are presented in Appendix~\ref{sec:VMC}.

\begin{figure}[t]
\includegraphics[width=0.9\columnwidth]{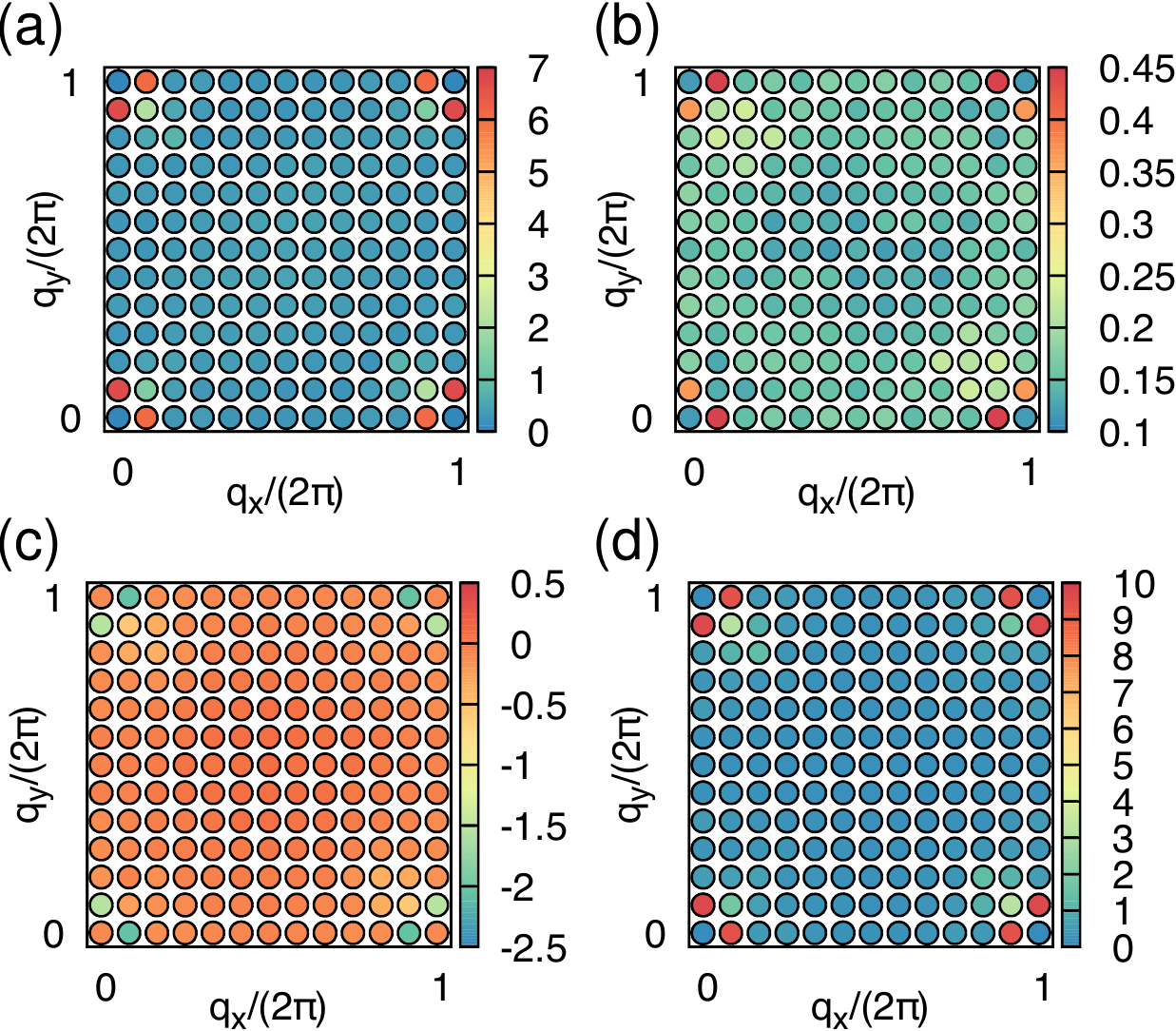}%
\caption{(Color online)
Charge structure factors at $(U/t,V/t)=(0,30)$ (phase separated
state) for the honeycomb lattice with $288$ sites, as obtained by 
VMC.
(a) Total charge structure factor $N(q)$.
(b) Charge-rich orbital ($c$-orbital-$c$-orbital) charge structure factor
$N^{cc}(q)$.
(c) Charge structure factor between two different orbitals ($c$-orbital
and $d$-orbital) $N^{cd}(q)$.
(d) Charge-poor orbital ($d$-orbital-$d$-orbital) charge structure factor
$N^{dd}(q)$.
The total charge structure factor $N(q)$ shows sharp peaks at the
achievable smallest wave vector $q$, suggesting phase separation.
Since $N^{dd}(q_{\rm peak})\gg N^{cc}(q_{\rm peak})$ and
$N(q)$ is similar to the $d$-orbital (charge-poor orbital) charge structure factor
$N^{dd}(q)$, phase separation mainly occurs in the $d$-orbital sector.
}
\label{fig:U0_VMC_Nq}
\end{figure}

\begin{figure*}[t]
\includegraphics[width=0.85\textwidth]{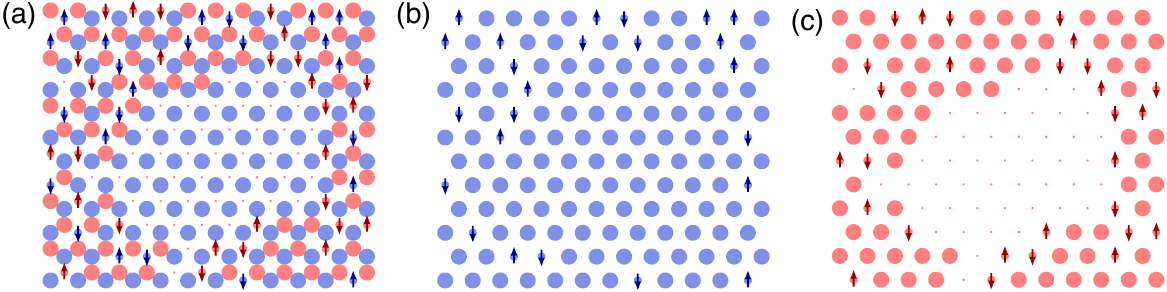}
\caption{(Color online)
Snapshot of a phase separated state
at $(U/t,V/t)=(0,30)$ for $288$ sites in the VMC calculation.
Large and small circles correspond to doublons and
spinons, respectively. Dots correspond to holons. Up and down arrows
correspond to up and down spins, respectively.
(a) Spin and charge configurations for both orbitals.
(b) Same snapshot for only charge-rich $c$ orbital.
It mainly consists of doublons.
(c) Same snapshot for only charge-poor $d$ orbital.
It consists of large doublon and holon islands.}
\label{fig:U0_VMC_snap_shot}
\end{figure*}

The Hartree-Fock calculation suggests that a triangular-like charge
order appears at large $V$.  We first check with variational Monte Carlo
(VMC) the stability when $U$ 
as well as $V$ are large. We confirm the presence of the insulating 
state with charge order and stripe antiferromagnetic order at 
$U/t=V/t=10$. Both the number of electrons per orbital and the magnetization 
are nearly saturated as shown in Figs.~\ref{fig:UV_VMC}(d)--\ref{fig:UV_VMC}(f).
The momentum distribution $n(k)$ is a smooth  function
of  $k$ [Fig.~\ref{fig:UV_VMC}(f)], suggesting the state to be insulating.
Note that our variational wave function also finds
a metallic state without charge and magnetic orders at $U/t=V/t=2$
[see Figs.~\ref{fig:UV_VMC}(a)--\ref{fig:UV_VMC}(c)]. Figure~\ref{fig:UV_PD}
presents the schematic phase diagram for $V=U$ obtained with the various
approaches considered here.

Besides, we do not find any indication of phase separation,
which can be detected by the divergence of the charge structure 
factor $N(q)$ at the smallest achievable wave vector 
$q\sim 2\pi/L$~\cite{becca2000}.

We now investigate the case of $U=0$ and large $V$, namely, 
$V/t=10$, $20$, $30$, with the VMC method. Note that perturbation theory is not 
applicable in this case since $U$ is not large enough. When 
$V/t=10$, the charge ordered metallic state found in the mean-field 
calculation is replaced by a metal without any charge order.
The total charge structure factor $N(q)$, see Eq.~(\ref{eq:def_structure_factor_Nq}), 
shows $q$-linear behavior near $q\sim 0$, suggesting the state
to be metallic.

On the other hand, when $V/t=20$ and $30$, we find a charge
disproportionate state,
where the average number of $c$ electrons is larger than that of $d$
electrons.
As shown in Fig.~\ref{fig:U0_VMC_Nq},
$N(q)$ is found to have sharp peaks near $q\sim 2\pi/L$, suggesting phase
separation~\cite{becca2000}.
The peaks in $N(q)$ appear to be dominated by that of the charge-poor
$d$-orbital charge structure factor $N^{dd}(q)$; see Eq.~(\ref{eq:def_structure_factor_Nabq}).
This means that phase separation is mainly activated in
the $d$-orbital sector.

In order to clarify the mechanism of  phase separation,
we also take a snapshot of this state.
As shown in Fig.~\ref{fig:U0_VMC_snap_shot}(a), phase 
separation is characterized by the charge ordered insulating state
with a 2020$\cdots$ structure (doubly occupied--empty--doubly occupied--empty $\cdots$
sites) and the metallic state with a mixture 
of doubly occupied and singly occupied sites.
Following the conventions~\cite{anderson1988},
we denote single-occupied sites with spin as ``spinons'' while
doubly occupied (empty) sites with no spin as ``doublons'' (``holons'').
 For the charge-rich $c$ orbital, each site 
is nearly doubly occupied and there are no holon sites (empty sites)
[see Fig.~\ref{fig:U0_VMC_snap_shot}(b)]. On the other hand, for 
the charge-poor $d$ orbital there are two islands: one formed by 
holons and the other one that is a mixture of doublons and spinons
[see Fig.~\ref{fig:U0_VMC_snap_shot}(c)].
In this doublon-spinon mixture region, each spinon can hop through a
doublon sea of the $c$ and $d$ orbitals. This does not cost an additional
energy if two spinons are not next to each other on the original
honeycomb lattice. Each spinon is always surrounded by three doublons,
which reside on the nearest-neighbor sites of the honeycomb lattice. It
can be assumed that one spinon and at least one doublon are bound
together, and this new quasiparticle freely moves inside the doublon
sea. The total kinetic energy gain is determined by the size of the
doublon sea and the effective filling of the new quasiparticles.
We note that the concept of spin-charge separation has been
only rigorously defined in one
spatial dimension; however, such a possibility has
been also discussed in higher dimensions in the presence of geometrical
frustrations~\cite{balents2010,poilblanc2009}.
Our
numerical results imply that the system wants to generate a larger
doublon sea with spinons to gain kinetic energy.
This mechanism is similar to what has been found in the doped
extended Hubbard model on a one-dimensional chain~\cite{mila1993,penc1994}
and a two-leg ladder~\cite{vojta2001}.

Summarizing, a nearest-neighbor Coulomb interaction $V$ at $U=0$ 
stabilizes a charge ordered metal at the Hartree-Fock level. 
However inclusion of quantum fluctuations via the 
Gutzwiller approximation (GA), see
Appendix~\ref{sec:Gutzwiller}, and via finite-size VMC
calculations suggests that the charge-ordered metal is replaced 
by phase separation, see Fig.~\ref{fig:U0_PD}, 
although the energies of these two states are found
to be very close, as shown in Appendix~\ref{sec:U=0}.
Figure~\ref{fig:U0_PD} shows the schematic phase diagram for $U=0$ and 
finite $V$. 
We note that the critical $U$ is shifted to a larger value in the VMC result.

\begin{figure}[t]
\includegraphics[width=0.9\columnwidth]{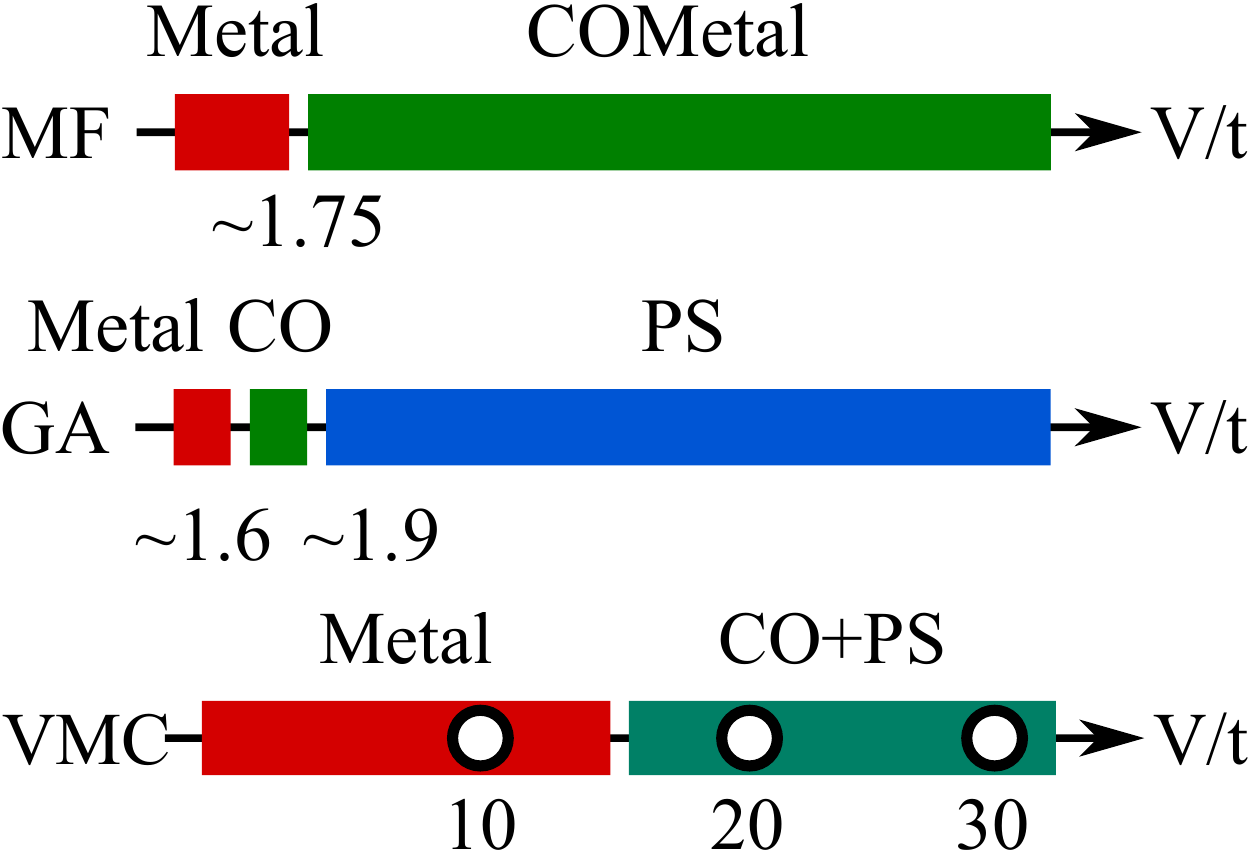}
\caption{(Color online)
Schematic phase diagrams for the honeycomb lattice at $3/4$ 
filling and $U=0$ obtained using restricted Hartree-Fock
(MF, top), the Gutzwiller approximation (GA, middle) and
variational Monte Carlo (VMC, bottom). CO denotes a charge 
ordered phase, while PS denotes a phase separated phase.
}
\label{fig:U0_PD}
\end{figure}

\section{Emergent kagome and chain structures on a triangular system}
\label{sec:triangular}

\begin{figure}[t]
\includegraphics[width=.7\columnwidth]{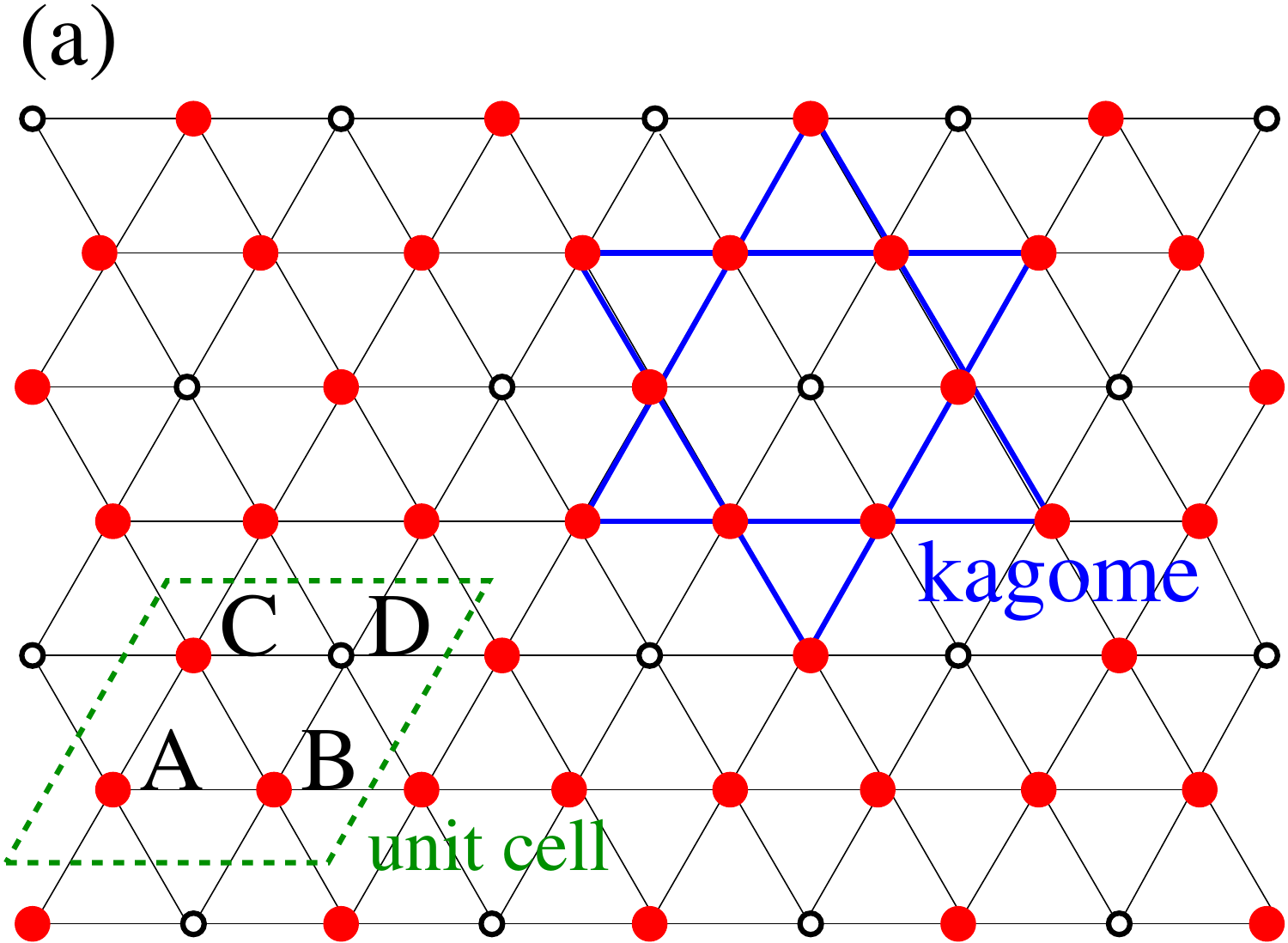}\\[\baselineskip]
\includegraphics[width=.7\columnwidth]{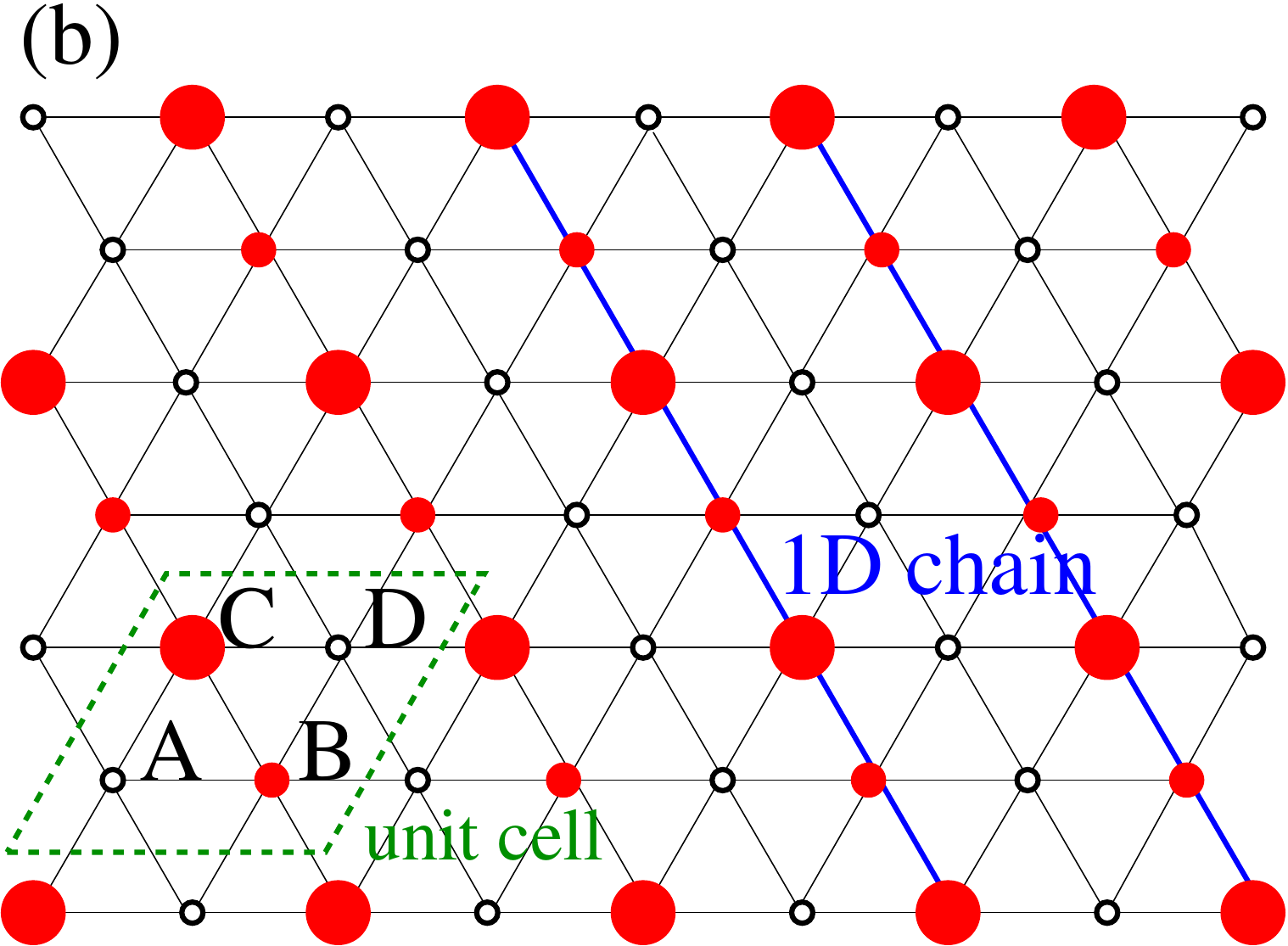}
\caption{(Color online) (a) Effective kagome lattice generated by 
charge order at $n=3/4$ on the triangular lattice. The unit cell contains 
four sites and is denoted by a green parallelogram. Small empty circles denote 
empty sites, while full red circles denote single-occupied sites. (b)
Effective 1D chains generated by charge order at $n=3/4$ on the triangular lattice. 
Small empty circles denote empty sites, full small red circles denote single-occupied sites, 
while full large red circles denote double-occupied sites. 
}
\label{fig:effective_lattices}
\end{figure}

 In this section we investigate the extended Hubbard model  
on the isotropic triangular lattice, as defined by the Hamiltonian:
\begin{equation}\label{eq:hubbard_triang}
\begin{split}
H=&-t\sum_{\langle i,j\rangle,\sigma} c^{\dagger}_{i,\sigma} c^{\phantom{\dagger}}_{j,\sigma} + \textrm{h.c.} +U\sum_i n_{i,\uparrow}n_{i,\downarrow} \\
& +V\sum_{\langle i,j\rangle} n_i n_j +V'\sum_{\langle\langle i, j\rangle \rangle} n_i n_j,
\end{split}
\end{equation}
where $t$ denotes the hopping parameter, $U$ is the on-site 
Coulomb repulsion, $V$ is the nearest-neighbor Coulomb interaction,
and $V'$ is the next-nearest-neighbor one. As in the previous 
section, we investigate repulsive Coulomb interactions, focusing 
on the appearance of charge-ordered states induced by a nonlocal potential.
Here, we focus on 3/8 filling ($n=3/4$), where emergent kagome 
and one-dimensional structures may be generated by the appearance 
of charge order. Both effective lattices are shown in Fig.~\ref{fig:effective_lattices}. 
When $U\gg V$, double occupancies are prohibited and the charge 
ordered ground state has a kagome-like structure, with three sites 
of the unit cell singly occupied and one site empty. By increasing 
the ratio $V/U$, the number of empty sites increases in order to 
avoid the energy loss from the $V$ term, thus inducing a 
one-dimensional (1D) charge structure. We remark that the presence 
of a nearest-neighbor Coulomb repulsion is not sufficient to 
stabilize the aforementioned charge orders and one additional 
next-nearest neighbor $V'$ is necessary in the triangular lattice 
case. Indeed, if we consider for example the kagome-like order of 
Fig.~\ref{fig:effective_lattices} (a), we can describe it as alternate 
rows that are fully occupied and rows where only half of the sites 
are occupied.  If interaction is restricted to nearest neighbors, 
the reciprocal positions of the empty sites between different rows 
can be changed without any further energy cost, implying the absence 
of charge order in the system.

In the $t\to 0$ limit, the energy 
of the kagome and the 1D phases can be easily computed being equal to 
$E=3/2(V+V')$ for the kagome substructure and to $E=V+V'+U/4$ for 
the 1D one. The 1D phase is then more favorable than the kagome 
one when $V+V'\ge U/2$. Here we set $U=30$, in analogy with our former 
investigation of charge ordered phases on the triangular 
lattice~\cite{tocchio2014}, and $V'=V/5$.

\begin{figure}[t]
\includegraphics[width=0.9\columnwidth]{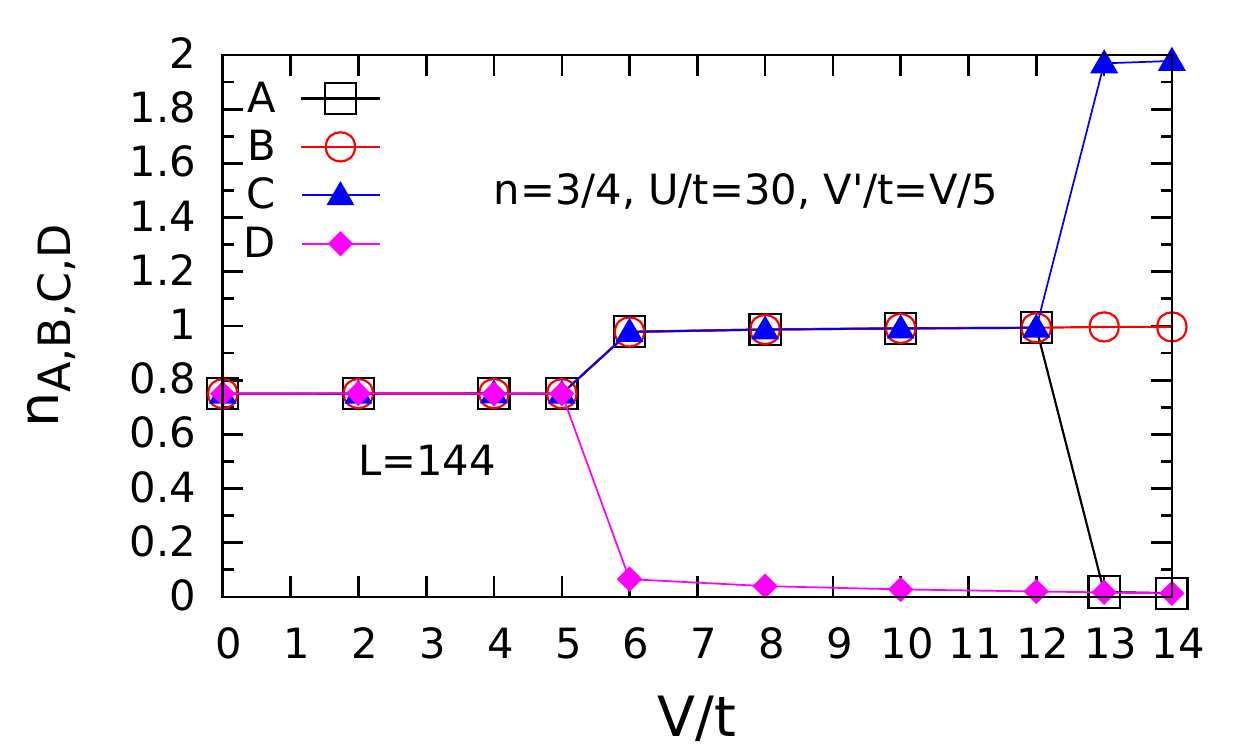}
\caption{(Color online)
 Electronic density $n_{\alpha}$
as a function of $V/t$ obtained from VMC calculations 
for each of the four sublattices $A$, $B$, $C$, and $D$, as 
present in the effective lattices emerging from the charge
ordered $n=3/4$ triangular lattice, as illustrated in 
Fig.~\ref{fig:effective_lattices}. The data are for $U/t=30$, 
$V'=V/5$, and a lattice size $L=144$.}
\label{fig:density}
\end{figure}

The model of Eq.~(\ref{eq:hubbard_triang}) is studied by means 
of the variational Monte Carlo method, the details being presented 
in Appendix~\ref{sec:VMC}. In order to distinguish the different 
kinds of charge ordering in the model, we plot in 
Fig.~\ref{fig:density} the average electronic density per 
sublattice $n_{\alpha}$, with $\alpha=A,B,C,D$ for each of the 
four sublattices that build up the unit cell (see 
Fig.~\ref{fig:effective_lattices}). Our results show that for 
$V/t \le 5$, the charge is uniformly distributed in the lattice, 
while for $6\le V/t\le 12$ one sublattice depletes, with the 
electrons forming an effective kagome lattice. In this case 
the frustration of the original lattice is effectively enhanced. 
Finally, as expected from the Coulomb energy argument, the 
1D substructure of Fig.~\ref{fig:effective_lattices} is 
stabilized for $V/t \ge 13$.

\begin{figure}[t]
\includegraphics[width=0.9\columnwidth]{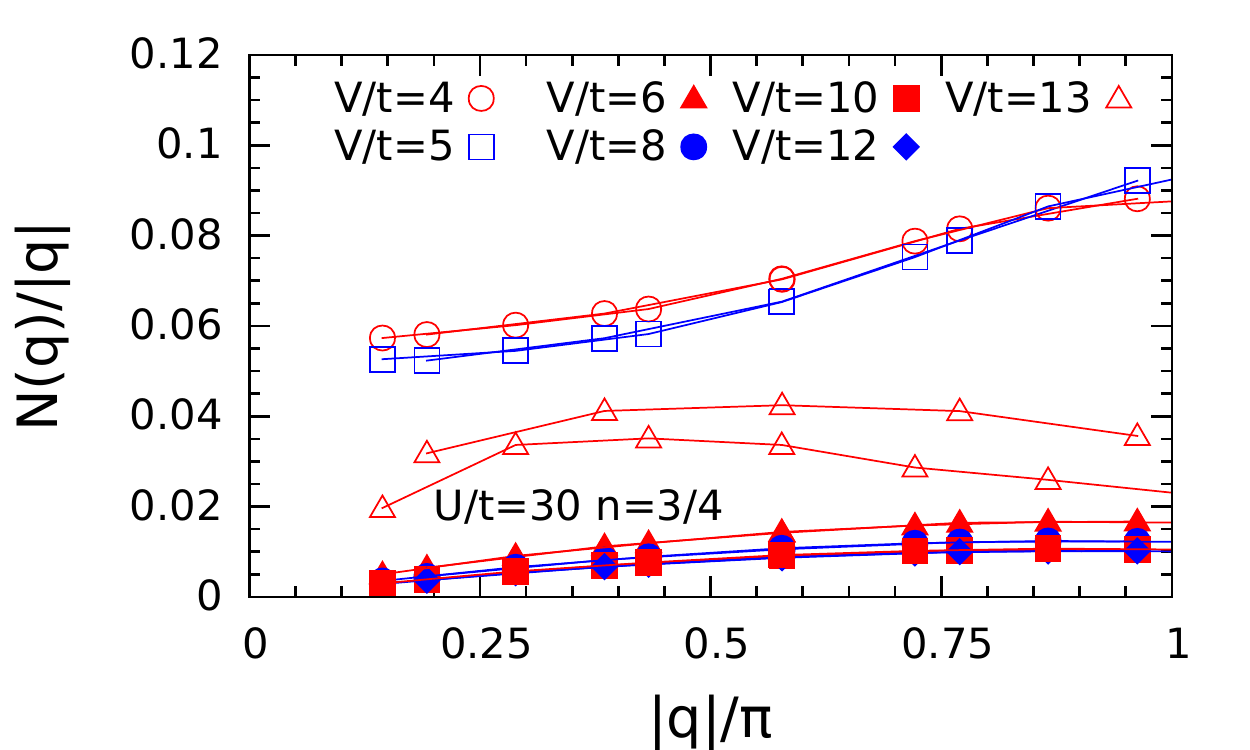}
\caption{(Color online) Variational Monte Carlo results for
$N(q)/|q|$ as a function of $|q|/\pi$ for
different values of $V/t$. The data are for $n=3/4$
and $U/t=30$ for the triangular lattice and for momenta $q$ connecting 
$\Gamma=(0,0)$ and $M=(\pi,\pi/\sqrt{3})$. The results for lattice 
sizes $L=144$ and $L=256$ are superimposed.} 
\label{fig:N_q}
\end{figure}

As discussed also in the honeycomb lattice section, the 
static structure factor $N(q)=\langle n_q n_{-q}\rangle$ 
is a good indicator for metallic behavior. The metallic 
phase is characterized by $N(q)\propto |q|$ for $q\to 0$, 
which implies a vanishing gap for particle-hole excitations. 
On the contrary, $N(q)\propto q^2$ for $q\to 0$, implies
a finite charge gap and insulating 
behavior~\cite{tocchio2011,tocchio2014}. The results shown 
in Fig.~\ref{fig:N_q} indicate that the system is metallic 
in the absence of charge order ($V/t=4,5$), while the charge 
ordered state with an effective kagome lattice exhibits an 
insulating behavior ($V/t=6,8,10,12$).  $N(q)$ 
is shown along the path in the Brillouin zone connecting 
the point $\Gamma=(0,0)$ to the point $M=(\pi,\pi/\sqrt{3})$ 
but a similar behavior can be obtained also along other directions. 
The results for the 1D charge ordered phase at $V/t=13$ 
indicate also an insulating behavior although we observe a
dependence  on the path chosen in the Brillouin zone, 
with strong finite-size effects. By increasing the 
lattice size up to $L=400$ we find, however, an insulating 
behavior along all the selected paths (not shown).

\begin{figure}[t]
\includegraphics[width=0.9\columnwidth]{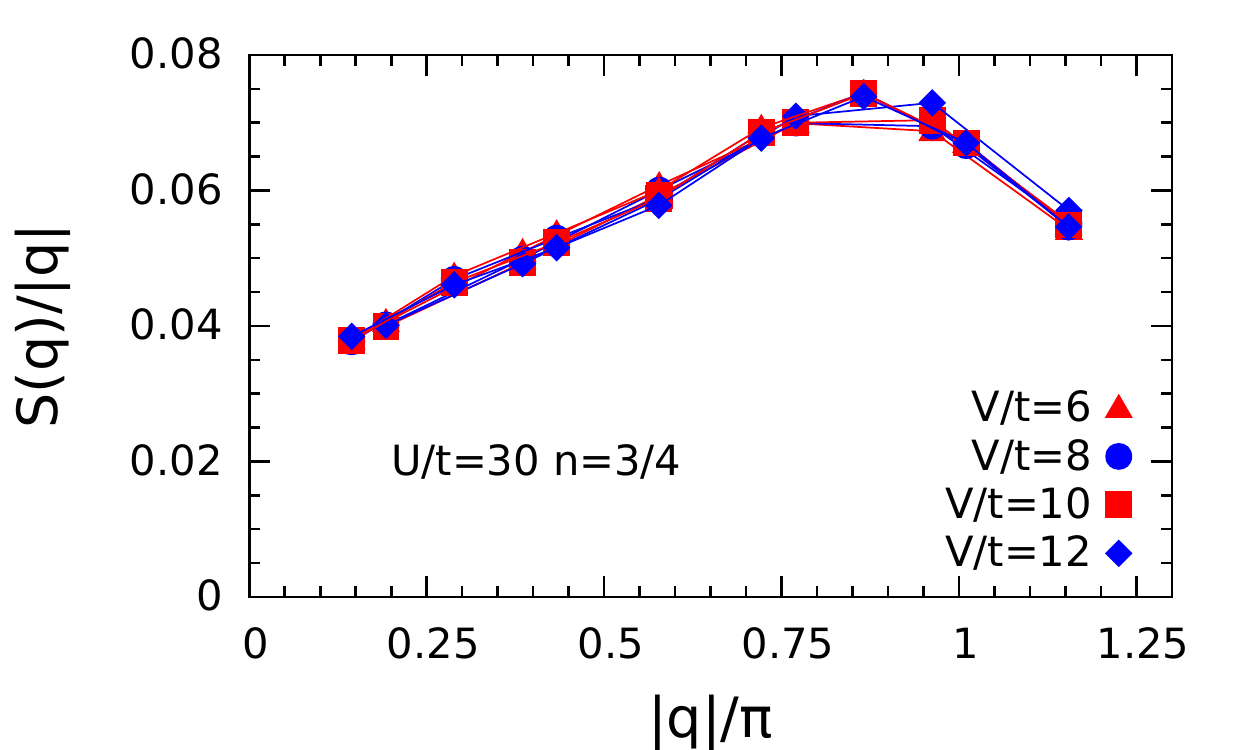}
\caption{(Color online) $S(q)/|q|$ as a function of $|q|/\pi$ for
different values of $V/t$, within the region where the effective kagome lattice is stabilized. 
Data are shown along the line between
$\Gamma=(0,0)$ and $M=(\pi,\pi/\sqrt{3})$ in the Brillouin zone on the $L=144$ 
and the $L=256$ lattice sizes.} 
\label{fig:S_q}
\end{figure}

In a similar way, one can consider the small-$q$ behavior of 
the spin-spin correlations $S(q)=\langle s_q s_{-q}\rangle$ 
to discriminate between a spin gapped and a spin gapless behavior. 
Our results indicate that the effective kagome lattice, induced 
by charge order, is characterized by gapless spin excitations, 
since $S(q)\propto |q|$ for $q\to 0$; see Fig.~\ref{fig:S_q}. 
Moreover, no peak can be observed in the spin-spin correlations, 
implying the absence of magnetic correlations, even at the 
short-range scale. We point out that gapless spin excitations have 
been also proposed for the Heisenberg model on the kagome lattice, 
by a similar variational approach~\cite{iqbal2013}, while the 
density matrix renormalization group approach suggests a 
finite gap in the spin excitations~\cite{depenbrock2012}.

\begin{figure}[t]
\includegraphics[width=0.9\columnwidth]{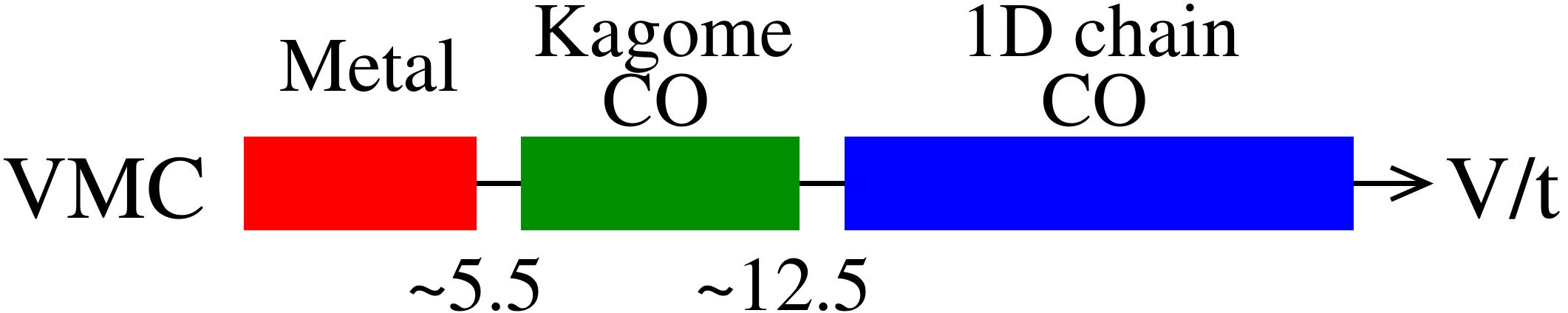}
\caption{(Color online) Schematic VMC phase diagram of the model of Eq.~(\ref{eq:hubbard_triang}) 
 as a function of $V/t$ at 3/8 filling, where we set $U/t=30$ and $V'=V/5$.
For $V/t \le 5$ we observe a metallic phase with a uniform charge distribution. For $6 \le V/t\le 12$ we stabilize the charge ordered insulator
with an effective kagome lattice of Fig.~\ref{fig:effective_lattices} (a). Finally, for $V/t \ge 13$, the charge ordered insulator
with effective 1D chains of Fig.~\ref{fig:effective_lattices} (b) occurs.} 
\label{fig:pd_triang}
\end{figure}

We finally summarize the VMC phase diagram of the model of Eq.~(\ref{eq:hubbard_triang}) 
at 3/8 filling, as a function of $V/t$, in Fig.~\ref{fig:pd_triang}.

\section{Conclusions}
\label{sec:conclusions}

In conclusion, by using a combination of Hartree-Fock, perturbation 
theory, and variational Monte Carlo, we have investigated the 
possibility of novel lattice structures emerging from charge 
disproportionation in doped systems via strong correlations.
In particular, we find an emergent geometrical frustration on 
bipartite honeycomb lattices, and an enhancement of the underlying 
geometrical frustration on a triangular lattice when Coulomb 
interactions beyond on-site are considered.

Concerning the honeycomb lattice, we have found that in the presence 
of both on-site  $U$ and nearest-neighbor $V$ Coulomb interactions, 
charge order converts the original honeycomb structure at $3/4$ filling
into an effective {\it half-filled} triangular lattice where the charge 
ordered state is characterized by a $2121\cdots$ ordered pattern,
while the singly occupied sites have a macroscopic spin degeneracy.
A nonzero hopping $t$ lifts the spin degeneracy by forming magnetic order,
which can be controlled by the Coulomb interactions $U$ and $V$.

Our analysis via Hartree-Fock of charge order and
spin correlations  shows
that most of the $U-V$ phase diagram at large values of $U$ and $V$
 is characterized by
a charge ordered antiferromagnetic insulator.
 This result is corroborated by VMC calculations 
for selected values of the parameter space.
The emergent antiferromagnetic spin correlations are consistent
with the effective antiferromagnetic Heisenberg model predicted
 by our perturbation theory analysis.
When $U$ is much larger than $V$, a charge ordered ferromagnetic insulating state
appears instead, which is consistent with the results
from  perturbation theory. By further decreasing $V$, charge
order completely disappears, and eventually a Nagaoka ferromagnetic semimetal~\cite{nagaoka1966}
appears for $V/t\gtrsim 6$. 

For $U=0$ and finite $V$, we find a charge ordered metal as the ground-state candidate.
Inclusion of quantum fluctuations via the VMC method, as well as the Gutzwiller
approximation, suggests however that this state may
be unstable towards phase separation
by forming a $2020\cdots$ charge ordered insulating region and a metallic one.

Concerning the triangular lattice, we find that effective kagome 
and one-dimensional lattices are generated at $3/8$ filling ($n=3/4$) because of the presence of charge order. 
We consider a large value of the on-site Coulomb repulsion $U$ and a small, but finite, value of the next-nearest-neighbor 
Coulomb interaction $V'=V/5$.
By increasing the ratio $V/U$ above $V/U\simeq 5.5$, the uniform metallic phase evolves into an insulating state, 
where the electronic charges form a kagome structure, each site being singly occupied. The emergence 
of a kagome lattice out of the original triangular one effectively enhances the frustration of the original lattice. 
The behavior of the spin-spin correlations $S(q)$ shows that the effective kagome lattice 
generated by charge order is nonmagnetic, with gapless spin excitations.
By further increasing the ratio $V/U$ above $V/U\simeq 12.5$, the number of empty sites 
increases in order to avoid the energy loss due to the $V$ term, thus generating another charge ordered insulator, 
where electrons form a one-dimensional charge structure.

Note added:
Recently we became aware of a paper~\cite{sugita2016} by Sugita and Motome
that reports the emergence of kagome and one-dimensional charge orders
on a triangular extended Hubbard model in the presence of spin-orbit coupling.

\acknowledgments

The authors would like to thank F. Becca, A. Kim, and S. M. Winter
for fruitful discussions.
R.K., R.V., and C.G. acknowledge the support of the German Science Foundation   
through Grant No.\ SFB/TRR49.
L.F.T. acknowledges the support of the Italian Ministry of Education, University,
and Research through Grant No. PRIN 2010 2010LLKJBX.
The variational Monte Carlo code,
which is used in the honeycomb lattice,
is based on a code first developed by
Tahara~\cite{tahara2008}.

\appendix
\counterwithin{figure}{section}
\counterwithin{table}{section}

\counterwithin{subsection}{section}
\makeatletter
\renewcommand{\p@subsection}{\thesection.\arabic{subsection}\expandafter\@gobble}
\makeatother

\section{Theoretical methods}
\label{sec:model_and_method}

For the honeycomb lattice, we have analyzed the extended Hubbard model on the honeycomb lattice by (1) the restricted Hartree-Fock method,
(2) the Gutzwiller approximation, (3) the 
variational Monte Carlo (VMC) method, and
(4) perturbation theory. For the extended Hubbard model (up to second neighbors) on the
triangular lattice we have restricted ourselves to the 
VMC
calculations. In this appendix
we show the details of the methods.

\subsection{Restricted Hartree-Fock method}
\label{sec:Hartree-Fock}

We consider a system size of
$N_{\mathrm{s}} = 2N_{\mathrm{dimer}} = 2L^2$.

The restricted Hartree-Fock approximation consists
of a mean-field decoupling of the Hamiltonian of Eq.~(\ref{eq:hubbard_2orbital}).
The mean-field Hamiltonian up to constant terms is given as
\begin{eqnarray}
\label{eq:meanfield}
 H_{\mathrm{MF}} &=&
  \sum_{\bm{k},\sigma}
  \left(
    \epsilon_{\bm{k}}c^{\dagger}_{\bm{k},\sigma} d_{\bm{k},\sigma}
  + \epsilon_{\bm{k}}^{*}d_{\bm{k},\sigma}^{\dagger} c_{\bm{k},\sigma}
  \right)
\nonumber
\\
 &&
  + U \sum_{i,\sigma}
  \left(
    \left<n_{i,\sigma}^{c}\right> n_{i,\bar\sigma}^{c}
  + \left<n_{i,\sigma}^{d}\right> n_{i,\bar\sigma}^{d}
  \right)
\nonumber
\\
 &&
  + V \sum_{i}
  \bigl(
    \left<n_{i}^{d}\right> n_{i}^{c}
  + \left<n_{i}^{c}\right> n_{i}^{d}
  + \left<n_{i}^{d}\right> n_{i+\bm{e}_x}^{c}
\nonumber
\\
 &&
  + \left<n_{i+\bm{e}_x}^{c}\right> n_{i}^{d}
  + \left<n_{i}^{d}\right> n_{i+\bm{e}_y}^{c}
  + \left<n_{i+\bm{e}_y}^{c}\right> n_{i}^{d}
  \bigr)
\nonumber
\\
 &&
  - V \sum_{i,\sigma}
  \bigl(
    \left<d_{i,\sigma}^{\dagger}c_{i,\sigma}\right> c_{i,\sigma}^{\dagger}d_{i,\sigma}
\nonumber
\\
 &&
  + \left<d_{i,\sigma}^{\dagger}c_{i+\bm{e}_x,\sigma}\right> c_{i+\bm{e}_x,\sigma}^{\dagger}d_{i,\sigma}
\nonumber
\\
 &&
  + \left<d_{i,\sigma}^{\dagger}c_{i+\bm{e}_y,\sigma}\right> c_{i+\bm{e}_y,\sigma}^{\dagger}d_{i,\sigma}
  + \mathrm{h.c.}
  \bigr),
\\
 \epsilon_{\bm{k}} &=& - t (1+e^{-ik_x}+e^{-ik_y}),
\\
 n_{i}^{\alpha} &=& n_{i,\uparrow}^{\alpha} + n_{i,\downarrow}^{\alpha} ~ (\alpha=c,d).
\end{eqnarray}
For a collinear state,
we assume
\begin{eqnarray}
 \left<n_{i,\sigma}^{c}\right> &=& n_{\sigma}^{c} + (-1)^{R_i} \delta n_{\sigma}^{c},
\\
 \left<n_{i,\sigma}^{d}\right> &=& n_{\sigma}^{d} + (-1)^{R_i} \delta n_{\sigma}^{d},
\\
 \left<d_{i,\sigma}^{\dagger}c_{i,\sigma}\right> &=& \chi_{\sigma} + (-1)^{R_i} \delta \chi_{\sigma},
\\
 \left<d_{i,\sigma}^{\dagger}c_{i+\bm{e}_x,\sigma}\right> &=&
 \left<d_{i,\sigma}^{\dagger}c_{i+\bm{e}_y,\sigma}\right> \\
 &=& \eta_{\sigma} + (-1)^{R_i} \delta \eta_{\sigma},
\end{eqnarray}
which contains $16$ independent parameters.
Here, $(-1)^{R_i} = e^{i\bm{Q}\cdot\bm{R}_i}$ with the momentum $\bm{Q}=(\pi,\pi)$ for a $2\times 2$ sublattice.
We can rewrite the Hamiltonian as
\begin{eqnarray}
\label{eq:meanfield2}
 &&
 H_{\mathrm{MF}} =
  \sum_{\bm{k},\sigma}^{\mathrm{RBZ}}
  \begin{pmatrix}
   c_{\bm{k},\sigma}^{\dagger} & c_{\bm{k}+\bm{Q},\sigma}^{\dagger} & d_{\bm{k},\sigma}^{\dagger} & d_{\bm{k}+\bm{Q},\sigma}^{\dagger}
  \end{pmatrix}
 \nonumber
\\
 &&
  \begin{pmatrix}
   O_{c,\sigma} & \Delta_{c,\sigma} & \tilde\epsilon_{\bm{k},\sigma} & Y_{\bm{k},\sigma} \\
   \Delta_{c,\sigma} & O_{c,\sigma} & Y_{\bm{k}+\bm{Q},\sigma} & \tilde\epsilon_{\bm{k}+\bm{Q},\sigma} \\
   \tilde\epsilon_{\bm{k},\sigma}^{*} & Y_{\bm{k}+\bm{Q},\sigma}^{*} & O_{d,\sigma} & \Delta_{d,\sigma} \\
   Y_{\bm{k},\sigma}^{*} & \tilde\epsilon_{\bm{k}+\bm{Q},\sigma}^{*} & \Delta_{d,\sigma} & O_{d,\sigma} \\
  \end{pmatrix}
  \begin{pmatrix}
   c_{\bm{k},\sigma} \\ c_{\bm{k}+\bm{Q},\sigma} \\ d_{\bm{k},\sigma} \\ d_{\bm{k}+\bm{Q},\sigma}
  \end{pmatrix}
\end{eqnarray}
with
\begin{eqnarray}
\label{eq:meanfield3}
 O_{c,\sigma} &=& Un_{\bar\sigma}^{c} + 3V (n_{\uparrow}^{d}+n_{\downarrow}^{d}),
\\
 O_{d,\sigma} &=& Un_{\bar\sigma}^{d} + 3V (n_{\uparrow}^{c}+n_{\downarrow}^{c}),
\\
 \Delta_{c,\sigma} &=& U\delta n_{\bar\sigma}^{c} - V (\delta n_{\uparrow}^{d}+\delta n_{\downarrow}^{d}),
\\
 \Delta_{d,\sigma} &=& U\delta n_{\bar\sigma}^{d} - V (\delta n_{\uparrow}^{c}+\delta n_{\downarrow}^{c}),
\\
 \tilde\epsilon_{\bm{k},\sigma} &=& \epsilon_{\bm{k}} - V [\chi_{\sigma} + \eta_{\sigma} (e^{-ik_x}+e^{-ik_y})],
\\
 Y_{\bm{k},\sigma} &=& - V [ \delta\chi_{\sigma} + \delta\eta_{\sigma} (e^{-ik_x}+e^{-ik_y})].
\end{eqnarray}
We further diagonalize the Hamiltonian and determine the above
 parameters self-consistently.
The self-consistent equations are given by
\begin{eqnarray}
 n_{\sigma}^{c}
 &=& \frac{1}{N} \sum_{k}^{\mathrm{RBZ}}
  \left[
   \left<c_{k,\sigma}^{\dagger} c_{k,\sigma}\right>
   + \left<c_{k+Q,\sigma}^{\dagger} c_{k+Q,\sigma}\right>
  \right],
\quad\quad
\\
 n_{\sigma}^{d}
 &=& \frac{1}{N} \sum_{k}^{\mathrm{RBZ}}
  \left[
   \left<d_{k,\sigma}^{\dagger} d_{k,\sigma}\right>
   + \left<d_{k+Q,\sigma}^{\dagger} d_{k+Q,\sigma}\right>
  \right],
\\
 \delta n_{\sigma}^{c}
 &=& \frac{1}{N} \sum_{k}^{\mathrm{RBZ}}
  \left[
   \left<c_{k,\sigma}^{\dagger} c_{k+Q,\sigma}\right>
   + \left<c_{k+Q,\sigma}^{\dagger} c_{k,\sigma}\right>
  \right],
\\
 \delta n_{\sigma}^{d}
 &=& \frac{1}{N} \sum_{k}^{\mathrm{RBZ}}
  \left[
   \left<d_{k,\sigma}^{\dagger} d_{k+Q,\sigma}\right>
   + \left<d_{k+Q,\sigma}^{\dagger} d_{k,\sigma}\right>
  \right],
\\
 \chi_{\sigma}
 &=& \frac{1}{N} \sum_{k}^{\mathrm{RBZ}}
  \left[
   \left<d_{k,\sigma}^{\dagger} c_{k,\sigma}\right>
   + \left<d_{k+Q,\sigma}^{\dagger} c_{k+Q,\sigma}\right>
  \right],
\\
 \delta\chi_{\sigma}
 &=& \frac{1}{N} \sum_{k}^{\mathrm{RBZ}}
  \left[
   \left<d_{k,\sigma}^{\dagger} c_{k+Q,\sigma}\right>
   + \left<d_{k+Q,\sigma}^{\dagger} c_{k,\sigma}\right>
  \right],
\\
 \eta_{\sigma}
 &=& \frac{1}{N} \sum_{k}^{\mathrm{RBZ}}
  \left[
   e^{ik_x} \left<d_{k,\sigma}^{\dagger} c_{k,\sigma}\right>\right.\\
   &-& \left. e^{ik_x} \left<d_{k+Q,\sigma}^{\dagger} c_{k+Q,\sigma}\right>
  \right],
\\
 \delta\eta_{\sigma}
 &=& \frac{1}{N} \sum_{k}^{\mathrm{RBZ}}
  \left[
   e^{ik_x} \left<d_{k+Q,\sigma}^{\dagger} c_{k,\sigma}\right> \right.\\
   &-& \left. e^{ik_x} \left<d_{k,\sigma}^{\dagger} c_{k+Q,\sigma}\right>
  \right],
\end{eqnarray}
with
$N=L^2$
being the number of $k$ points in the
Brillouin zone.
The ground states are obtained by filling the lowest-energy orbitals
up to the number of electrons.
We employ an antiperiodic-periodic boundary condition, and set the number of $k$ points as $120\times120$.

Each state is characterized by the magnetization and the number of electrons
for orbitals $c$ and $d$, which are defined as
\begin{eqnarray}
\label{eq:def_order_magnetization}
 \left<S^{z,\alpha}\right> &=& \frac{1}{N_{\mathrm{dimer}}}
  \sum_{i} (-1)^{R_i} \left<S_{i}^{z,\alpha}\right>,
\\
\label{eq:def_order_charge}
 \left<n^\alpha\right> &=& \frac{1}{N_{\mathrm{dimer}}} 
  \sum_{i} \left<n_{i}^\alpha\right> 
\quad 
 (\alpha=c,d).
\end{eqnarray}

Similarly, for spiral states, we consider momentum $\bm{Q}=(-2\pi/3,2\pi/3)$ for a
$6$-site sublattice, namely, the order parameters are given as
$\langle n_{i,\sigma}^{c} \rangle = n_c$,
$\langle n_{i,\sigma}^{d} \rangle = n_d (< n_c)$,
$\langle c^{\dagger}_{i,\uparrow} c^{\dagger}_{i,\downarrow} \rangle =
m_c e^{i\bm{Q}\cdot\bm{R}_i}$ with $m_c=0$ (for charge-rich sites),
$\langle d^{\dagger}_{i,\uparrow} d^{\dagger}_{i,\downarrow} \rangle =
m_d e^{i\bm{Q}\cdot\bm{R}_i}$ with $m_d\not=0$ (for charge-poor sites),
$\langle d^{\dagger}_{i,\sigma} c^{\dagger}_{i,\sigma} \rangle = \chi$,
and
$\langle d^{\dagger}_{i,\sigma} c^{\dagger}_{i+\bm{e}_x(\bm{e}_y),\sigma} \rangle = \eta$,
and the corresponding reduced Brillouin zone.

We also calculate the gap in the density
of states at the Fermi level  to check whether the state is metallic or insulating.

\subsection{Gutzwiller approximation}
\label{sec:Gutzwiller}

We applied the Gutzwiller approximation to the extended Hubbard
model on the honeycomb lattice. 
In the Gutzwiller approximation, the nearest-neighbor correlations
$\langle T_{cd\sigma} \rangle = \langle c_{i,\sigma}^{\dagger} d_{j,\sigma} \rangle$ 
in the correlated wave function $|\psi\rangle$
are evaluated via those in the uncorrelated wave function $|\psi\rangle_0$
through the renormalization factor $\eta_{cd\sigma}$~\cite{ogawa1975,zhang1988}:
\begin{equation}
 \langle c_{i,\sigma}^{\dagger} d_{j,\sigma} \rangle
 = \eta_{cd\sigma} \langle c_{i,\sigma}^{\dagger} d_{j,\sigma} \rangle_0.
\end{equation}
The matrix elements of $c_{i,\sigma}^{\dagger} d_{j,\sigma}$
in the uncorrelated wave function $|\psi\rangle_0$
are proportional to
\begin{equation}
 \sqrt{n_{c\sigma}^{0}(1-n_{c\sigma}^{0})}
 \sqrt{n_{d\sigma}^{0}(1-n_{d\sigma}^{0})}.
\end{equation}
Here, $n_{c\sigma}^{0} = n_{d\sigma}^{0} = 3/4$ at $3/4$ filling.
Generalizing this to the matrix elements of the correlated state, we find
\begin{equation}
 \eta_{cd\sigma} =
 \frac
 {(\sqrt{h_{c}s_{c\sigma}}+\sqrt{s_{c\bar{\sigma}}d_{c}})
  (\sqrt{s_{d\sigma}h_{d}}+\sqrt{d_{d}s_{d\bar{\sigma}}})}
 {\sqrt{n_{c\sigma}^{0}(1-n_{c\sigma}^{0})}
  \sqrt{n_{d\sigma}^{0}(1-n_{d\sigma}^{0})}}.
\end{equation}
for the Gutzwiller approximation of the kinetic energy term $T_{cd\sigma}$.
We define occupancies for empty ($h$), singly ($s$), and doubly ($d$)
occupied sites for orbital $c$ (similarly for orbital $d$):
\begin{eqnarray}
 h_{c} &=& \langle (1-c_{\uparrow}^{\dagger}c_{\uparrow})
  (1-c_{\downarrow}^{\dagger}c_{\downarrow}) \rangle,
\\
 s_{c\sigma} &=& \langle c_{\sigma}^{\dagger}c_{\sigma}
  (1-c_{\bar{\sigma}}^{\dagger}c_{\bar{\sigma}}) \rangle,
\\
 d_{c} &=& \langle c_{\uparrow}^{\dagger}c_{\uparrow}
  c_{\downarrow}^{\dagger}c_{\downarrow} \rangle.
\end{eqnarray}
The expectation value of the nearest-neighbor Coulomb interaction is given as
\begin{equation}
 \langle (c_{\uparrow}^{\dagger}c_{\uparrow} + c_{\downarrow}^{\dagger}c_{\downarrow})
 (d_{\uparrow}^{\dagger}d_{\uparrow} + d_{\downarrow}^{\dagger}d_{\downarrow}) \rangle
 =
 (n_{c\uparrow} + n_{c\downarrow})
 (n_{d\uparrow} + n_{d\downarrow}).
\end{equation}

The energy per bond in the Gutzwiller approximation is finally given as
\begin{eqnarray}
\label{eq:GA_energy}
 E &=&
 - t \sum_{\sigma} \eta_{cd\sigma} \langle T_{cd\sigma} + T_{dc\sigma} \rangle_0
\nonumber
\\
 &&
 + \frac{U}{3} (d_c + d_d)
 + V (n_{c\uparrow} + n_{c\downarrow}) (n_{d\uparrow} + n_{d\downarrow}).
\quad\quad
\end{eqnarray}
The  $1/3$ factor in the $U$ term comes from the number of nearest-neighbor bonds per site.
Here,
the expectation value of the hopping is a function of
$n_{c\sigma} + n_{d\sigma}$, and is defined as
\begin{equation}
\label{eq:GA_hopping}
 - t\left<T_{cd\sigma}+T_{dc\sigma}\right>_0
 = \frac{1}{N_{\rm bond}} \sum_{|k|<k_F^{\sigma}} \epsilon_k
\end{equation}
with $N_{\rm bond} = 3L^2$.
Besides,
the renormalization factor $\eta_{cd\sigma}$ is a function of $h_{c(d)}$,
$s_{c(d)\sigma}$, and $d_{c(d)}$.
One can eliminate the spinon $s_{c\sigma}$ and the holon $h_{c}$ using
\begin{eqnarray}
 s_{c\sigma} &=& n_{c\sigma} - d_{c},
\\
 h_{c} &=& 1 - s_{c\uparrow} - s_{c\downarrow} - d_{c}
  = 1 - n_{c\uparrow} - n_{c\downarrow} + d_{c}.
\quad\quad
\end{eqnarray}
Therefore, the energy becomes a function of the parameters
$n_{c\sigma}$,
$n_{d\sigma}$,
$d_{c}$,
and
$d_{d}$.

Naively, the energy minimum can be obtained by the condition
\begin{equation}
 \frac{\partial E}{\partial d_{c}}
 = \frac{\partial E}{\partial d_{d}}
 = 0.
\end{equation}
In the absence of on-site Coulomb interaction ($U=0$), this yields
\begin{eqnarray}
\label{eq:GA_V0_condition_d=nn}
 d_{c} &=& n_{c\uparrow} n_{c\downarrow},
\\
 d_{d} &=& n_{d\uparrow} n_{d\downarrow}.
\end{eqnarray}
The expectation value of the doublon is a simple product of the number of up and down spins.
This simplifies the renormalization factor:
\begin{equation}
\label{eq:GA_V0_condition_eta}
 \eta_{cd\sigma} =
 \frac
 {\sqrt{n_{c\sigma}(1-n_{c\sigma})}
  \sqrt{n_{d\sigma}(1-n_{d\sigma})}}
 {\sqrt{n_{c\sigma}^{0}(1-n_{c\sigma}^{0})}
  \sqrt{n_{d\sigma}^{0}(1-n_{d\sigma}^{0})}}.
\end{equation}
Now, the energy is a function of a few parameters, namely, 
$n_{c\sigma}$ and $n_{d\sigma}$. The energy minimum can 
be searched analytically. For $U\not=0$, however, the 
stationary condition does not give us simple conditions.
We, instead, numerically find the energy minimum by 
controlling parameters
$n_{c\sigma}$, $n_{d\sigma}$, $d_{c}$, and $d_{d}$.

\begin{figure}[t]
\includegraphics[width=0.8\columnwidth]{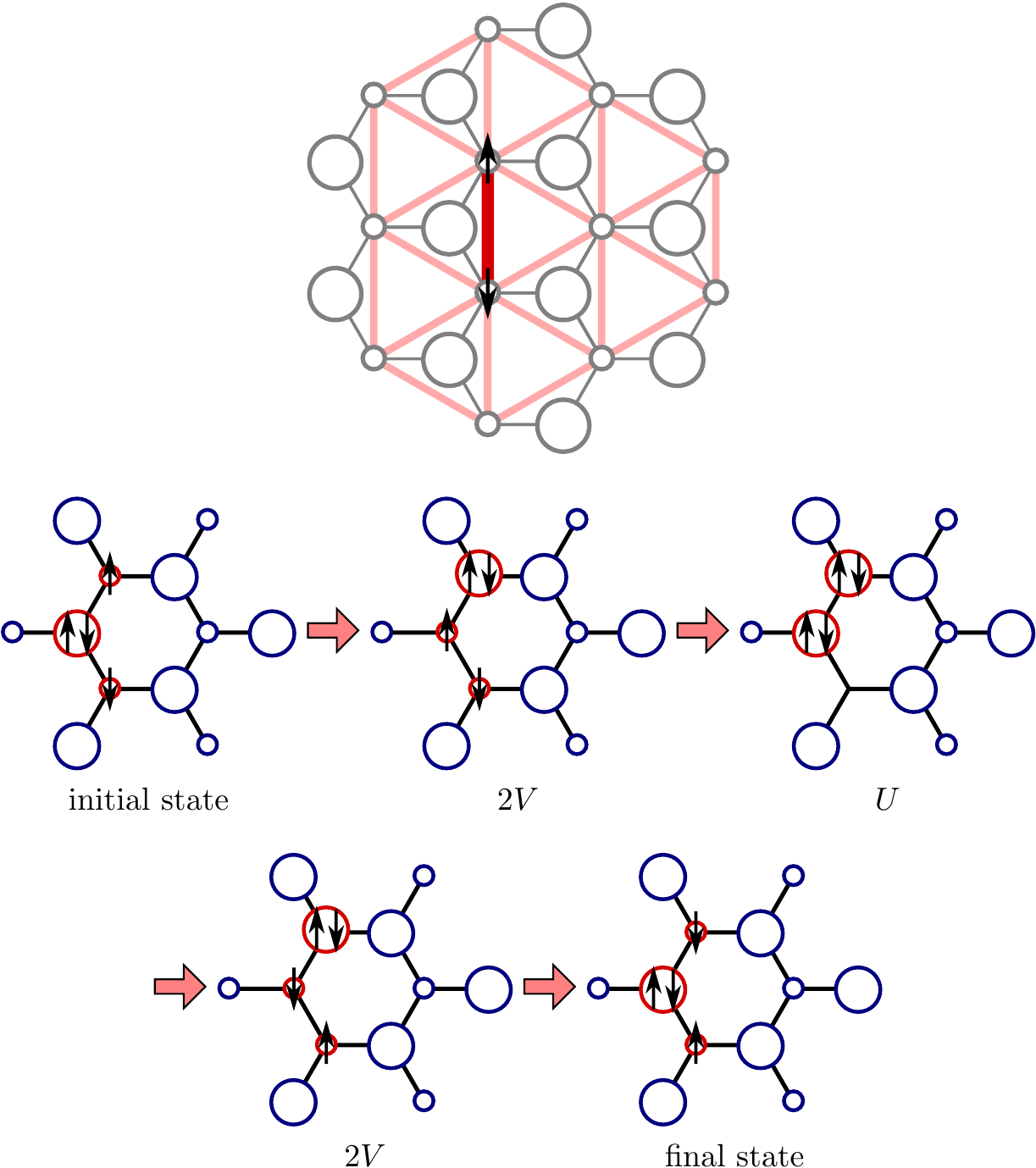}%
\caption{(Color online)
Virtual hopping process which generates an effective 
nearest-neighbor spin exchange interaction between the 
triangular lattice sites emerging from the charge ordered 
$n=3/2$ 
honeycomb
lattice. Compare Eq.~(\ref{eq_A_H_spin}).}
\label{fig:UV_Jeff_Jnn}
\end{figure}

\subsection{Variational Monte Carlo method}
\label{sec:VMC}

\begin{figure}[t]
\includegraphics[width=0.8\columnwidth]{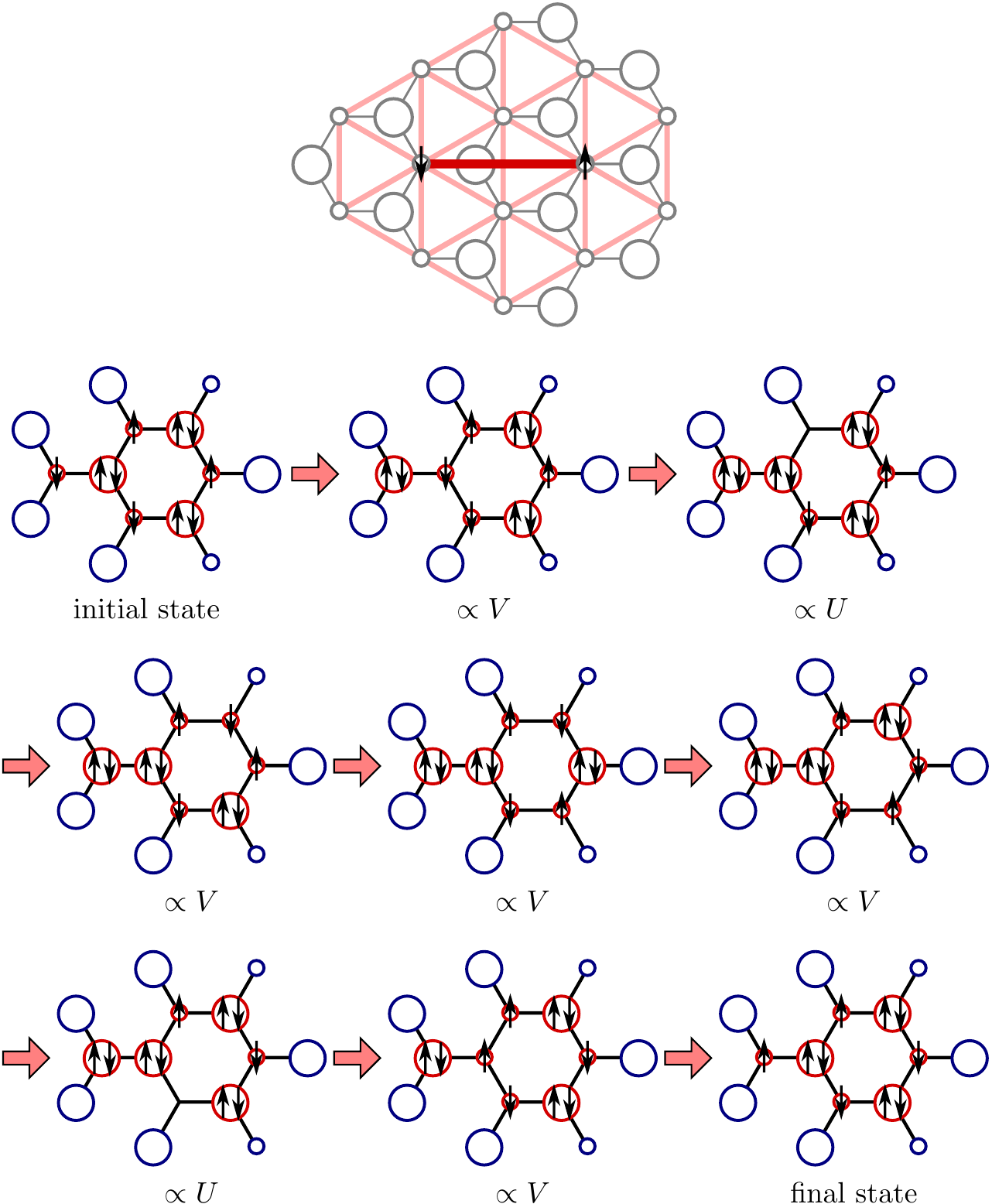}%
\caption{(Color online)
Virtual hopping process which generates an effective 
next-nearest-neighbor spin exchange interaction between the 
triangular lattice sites emerging from the charge ordered 
$n=3/2$ 
honeycomb
lattice. Compare Eq.~(\ref{eq_A_H_spin}).}
\label{fig:UV_Jeff_Jnnn}
\end{figure}

As a third method which includes the effects of quantum
fluctuations beyond mean field, we consider the variational
Monte Carlo (VMC) technique.
We use the Jastrow-Slater  wave functions which allow
metallic and insulating states with charge and antiferromagnetic
orders. For the honeycomb lattice Eq.~(\ref{eq:hubbard_2orbital}), we define
\begin{eqnarray}
 \left|\psi\right> &=&
  P_{\mathrm{CJ}} P_{\mathrm{SJ}} \left|\phi\right>,
\\
 \left|\phi\right> &=&
  \biggl[
   \sum_{ij} \bigl(
    f_{ij}^{cc} c_{i,\uparrow}^\dagger c_{j,\downarrow}^\dagger
    + f_{ij}^{cd} c_{i,\uparrow}^\dagger d_{j,\downarrow}^\dagger
\nonumber
\\
 &&
    + f_{ij}^{dc} d_{i,\uparrow}^\dagger c_{j,\downarrow}^\dagger
    + f_{ij}^{dd} d_{i,\uparrow}^\dagger d_{j,\downarrow}^\dagger
   \bigr)
  \biggr]^{N_{\mathrm{e}}/2} \left|0\right>,
\\
 f_{ij} &=& \left\{
  \begin{array}{cc}
   f^{A}(\bm{r}_j-\bm{r}_i) & i \in \mathrm{A~sublattice}, \\
   f^{B}(\bm{r}_j-\bm{r}_i) & i \in \mathrm{B~sublattice},
  \end{array} \right.
\\
 P_{\mathrm{CJ}} &=&
  \exp\biggl[
   \frac{1}{2} \sum_{ij} \bigl(
    v^{\mathrm{CJ},cc}_{ij} n_{i}^{c} n_{j}^{c}
\nonumber
\\
 &&
    + v^{\mathrm{CJ},cd}_{ij} n_{i}^{c} n_{j}^{d}
    + v^{\mathrm{CJ},dd}_{ij} n_{i}^{d} n_{j}^{d}
   \bigr)
  \biggr],
\\
 P_{\mathrm{SJ}} &=&
  \exp\biggl[
   2 \sideset{}{'}\sum_{ij} \bigl(
    v^{\mathrm{SJ},cc}_{ij} S_{i}^{z,c} S_{j}^{z,c}
\nonumber
\\
 &&
    + v^{\mathrm{SJ},cd}_{ij} S_{i}^{z,c} S_{j}^{z,d}
    + v^{\mathrm{SJ},dd}_{ij} S_{i}^{z,d} S_{j}^{z,d}
   \bigr)
  \biggr],
\\
 v^{\mathrm{SJ}}_{ij} &=&
  v^{\mathrm{SJ}}(|\bm{r}_j-\bm{r}_i|),
\\
 v^{\mathrm{CJ}}_{ij} &=&
  v^{\mathrm{CJ}}(|\bm{r}_j-\bm{r}_i|).
\end{eqnarray}
Here, $\sum'_{ij}$ denotes a sum over $i\not=j$.
We prepare the Slater part $|\phi\rangle$ by taking the Hartree-Fock solutions
as initial states, and optimize the variational parameters $f_{ij}$,
$v^{\mathrm{SJ}}_{ij}$, and $v^{\mathrm{CJ}}_{ij}$.

\begin{figure}[t]
\includegraphics[width=0.8\columnwidth]{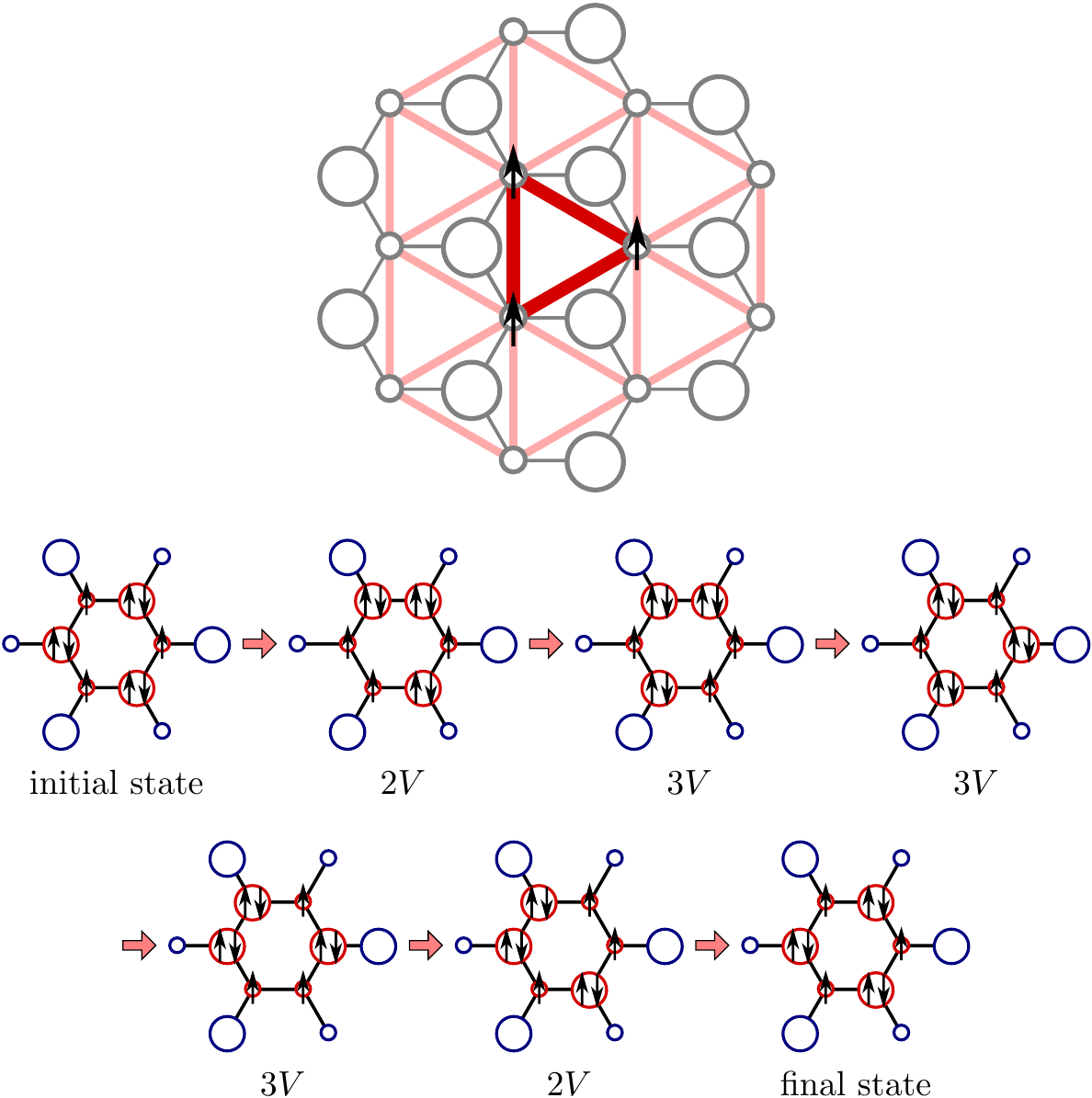}%
\caption{(Color online)
Virtual hopping process which generates an effective 
three-spin permutation $K_3^{\triangleright}$ (surviving
even in the limit $U\to\infty$) between the triangular 
lattice sites emerging from the charge ordered
$n=3/2$
honeycomb
lattice. Compare Eq.~(\ref{eq_A_H_spin}).}
\label{fig:UV_Jeff_Jr3}
\end{figure}

We use Eqs.~(\ref{eq:def_order_magnetization}) and
(\ref{eq:def_order_charge}) to characterize each phase.
In order to see whether the phase is metallic or insulating,
we calculate the total momentum distribution:
\begin{equation}
\label{eq:def_mom_dist_nk}
 n(k) = \frac{1}{2N_{\mathrm{s}}}\sum_{ij\sigma}
 \left<
   c_{i,\sigma}^{\dagger} c_{j,\sigma}
 + d_{i,\sigma}^{\dagger} d_{j,\sigma}
 \right>
 e^{ik\cdot(r_i-r_j)},
\end{equation} 
and the density-density structure factors for two orbitals:
\begin{equation}
\label{eq:def_structure_factor_Nabq}
 N^{\alpha\beta}(q) =
  \frac{1}{N_{\mathrm{dimer}}}\sum_{i,j}
  \left<n_i^\alpha n_j^\beta\right>
  e^{iq\cdot(r_i-r_j)}
 \quad (\alpha=c,d).
\end{equation}
Metallic states are detected by the jump of the momentum distribution $n(k)$
and $q$-linear behavior of the total charge structure factor
\begin{equation}
\label{eq:def_structure_factor_Nq}
 N(q) = N^{cc}(q) + N^{cd}(q) + N^{dc}(q) + N^{dd}(q)
\end{equation}
near $q\sim 0$. On the other hand, $n(k)$ is smooth and
$N(q)\sim q^2$ ($q\sim 0$) for insulating states.

Analogously, in order to simulate the triangular
lattice model of Eq.~(\ref{eq:hubbard_triang}), 
we have used the variational Monte Carlo method based on the 
variational ansatz 
$|\Psi\rangle=P_{\textrm{CJ}}|\textrm{FS}\rangle$~\cite{gros1988,zhang1988,capello2005}, 
where $|\textrm{FS}\rangle$ is the noninteracting filled Fermi sea, 
to which a finite small superconductive term is added in order to 
regularize the wave function, i.e., to separate the highest occupied 
and the lowest unoccupied states by a gap. The term 
\begin{equation}
P_{\textrm{CJ}}=\exp\left({-\frac{1}{2}\sum_{i,j}v_{ij}n_in_j}\right)
\end{equation}
is the density-density Jastrow factor~\cite{capello2005}, where 
the $v_{ij}$'s are optimized
with variational Monte Carlo calculations for every
independent distance $|\bm{r}_i-\bm{r}_j|$ (including on site). 
In order to investigate the formation of charge-ordered phases, 
we include four different chemical potentials in $|\textrm{FS}\rangle$, 
as variational parameters, one
for each site of the unit cell, similarly to what has been done 
in Ref.~\onlinecite{tocchio2014}. We tested that
inclusion of backflow correlations to further
improve the correlated 
state~$|\Psi\rangle$~\cite{tocchio2008,tocchio2011} 
is not crucial to describe charge-ordered states. 
All results presented here are obtained by optimizing individually~\cite{yunoki2006} 
every variational parameter in the wave function and then performing a 
Monte Carlo sampling of the observables over the optimal state. The 
error bars are not shown since they are always smaller than the symbol size.

\subsection{Details of the perturbation calculations on honeycomb systems}
\label{sec:Perturbation}

\begin{figure}[t]
\includegraphics[width=0.8\columnwidth]{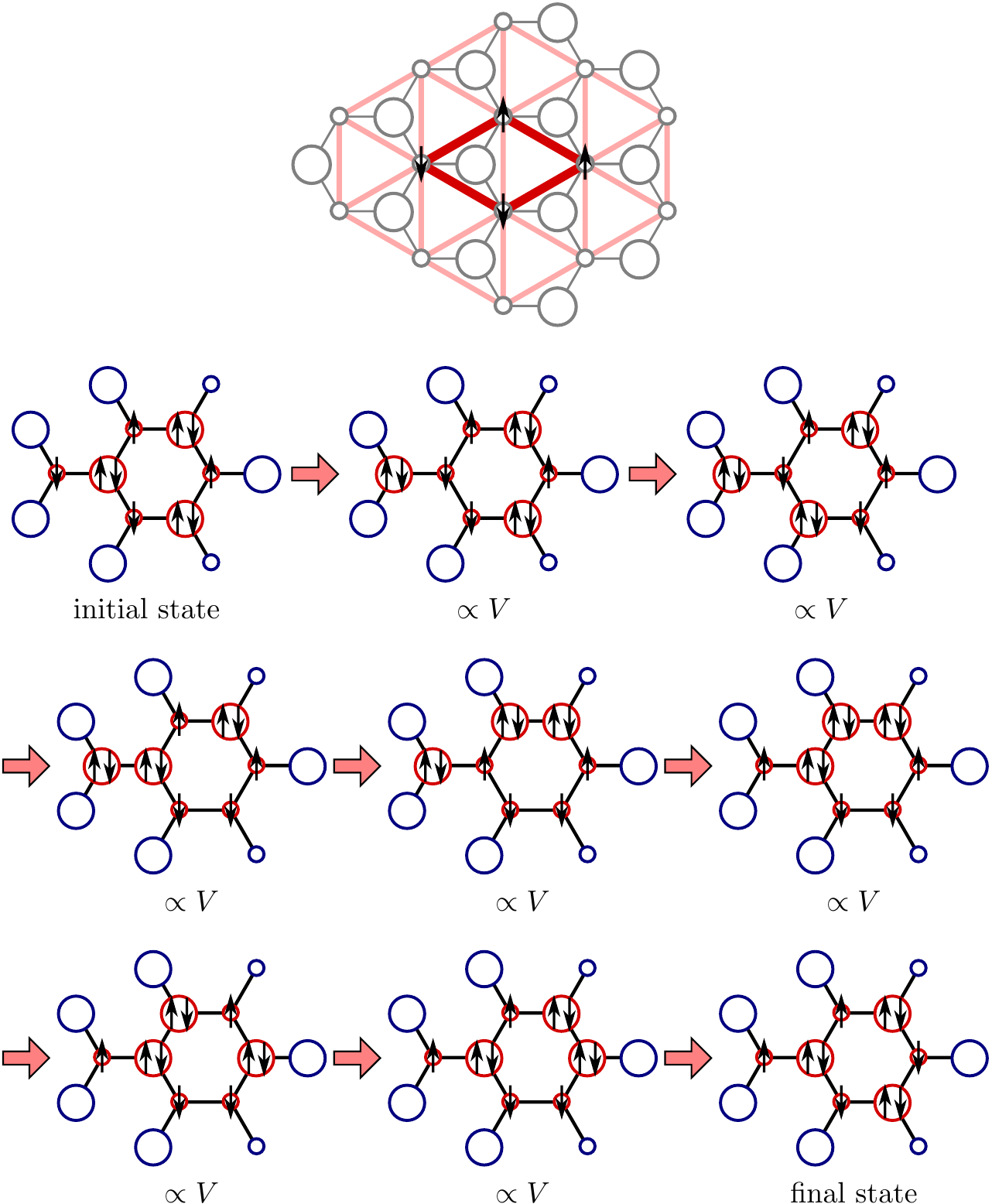}%
\caption{(Color online)
Virtual hopping process which generates an
effective four-spin permutation between the triangular 
lattice sites emerging from the charge ordered
$n=3/2$
honeycomb
lattice. Compare Eq.~(\ref{eq_A_H_spin}).}
\label{fig:UV_Jeff_Jr4}
\end{figure}

On the honeycomb lattice, when $U=V=\infty$,
we expect a triangular charge order at $3/4$ filling ($n=3/2$).
Charge-rich sites contain two electrons per site, and do not have
any left spin degrees of freedom.
On the other hand, charge-poor sites contain one electron per site,
and possess macroscopic spin degeneracy.
The ground-state degeneracy is lifted in the presence of hopping $t$.
From  perturbation theory,
the effective Hamiltonian can be obtained as a sum of
the Heisenberg spin exchange  and  permutation terms:
\begin{eqnarray}
 H
 &=& \sum_{\langle i,j\rangle_1} J_1 \left(\bm{S}_i\cdot\bm{S}_j - \frac{1}{4}\right)
\nonumber
\\
 &&
 + \sum_{\langle i,j\rangle_2} J_2 \left(\bm{S}_i\cdot\bm{S}_j - \frac{1}{4}\right)
 + \cdots
\nonumber
\\
 &&
 + \sum_{\vartriangleright} K_{3}^{\triangleright} (P_{3}+P_{3}^{-1})
\nonumber
\\
 &&
 + \sum_{\vartriangleleft} K_{3}^{\triangleleft} (P_{3}+P_{3}^{-1})
\nonumber
\\
 &&
 + \sum_{\square} K_{4} (P_{4}+P_{4}^{-1})
 + \cdots.
\label{eq_A_H_spin}
\end{eqnarray}
For the spin exchange terms the sum is taken over all 
nearest-neighbor (next-nearest-neighbor) sites on an 
effective triangular lattice for 
$\langle i,j\rangle_1$ ($\langle i,j\rangle_2$).
On the other hand, for the permutation terms,
the sum is taken over all right-pointing
(left-pointing) triangles which are located 
inside (outside) of hexagons for 
$\vartriangleright$ ($\vartriangleleft$)
and all squares for $\square$.
The symbol $P_n$ denotes a cyclic permutation operator.

\begin{figure}[t]
\includegraphics[width=0.9\columnwidth]{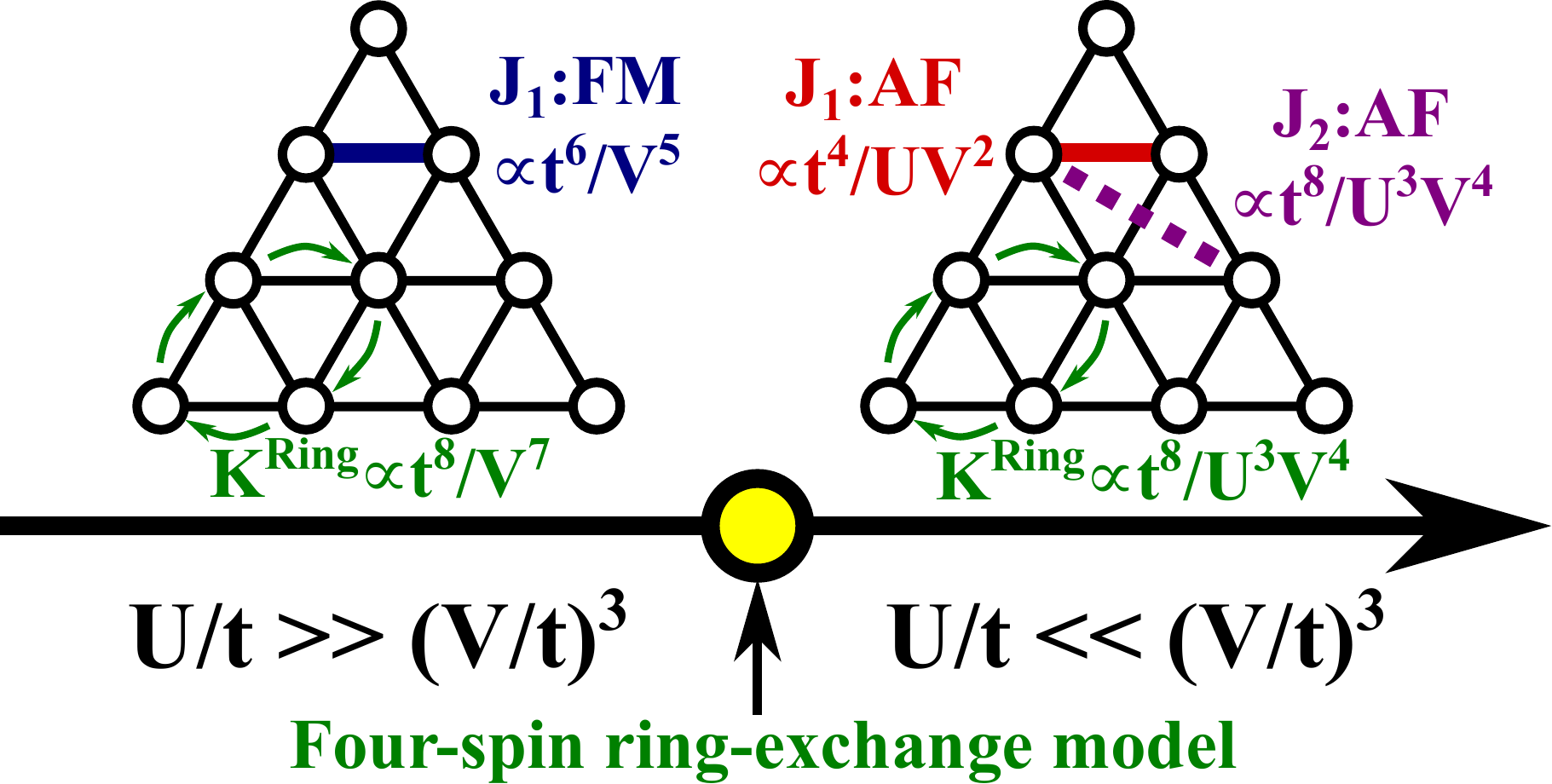}
\caption{(Color online)
Effective Hamiltonians for the triangular lattice emerging
from the charge ordered $n=3/2$
honeycomb
lattice, see
Eq.~(\ref{eq_A_J_eff}), as a function of $U/t$ and $V/t$.}
\label{fig:UV_Jeff_PD}
\end{figure}

Unlike the perturbation expansion in the triangular and kagome systems
where electrons can hop odd times in local triangles, spin exchange 
interactions appear through only an even number of electron hoppings.
Besides, the spin degeneracy is first lifted by a four virtual hopping process
rather than a conventional two hopping process.
For example, as shown in Fig.~\ref{fig:UV_Jeff_Jnn},
nearest-neighbor exchange on the triangular lattice appears
through four hopping processes.
The intermediate states have energy $2V$, $U$, and $2V$, respectively.
There are four different ways of exchanging spins.
This gives antiferromagnetic exchange
$J_1 = 4 t^{4}/[(2V)^2U] = t^{4}/(V^2U)$.
Similarly, next-nearest-neighbor exchange on the triangular lattice
is given as 
$J_2 = c_2 t^{8}/(V^5U^2) + c'_2 t^{8}/(V^4U^3)$ with $c_2$, $c'_2$
being positive constants (see Fig.~\ref{fig:UV_Jeff_Jnnn}).
These exchange interactions are always antiferromagnetic:
\begin{equation}
J \propto \frac{t^{2n}}{\mathrm{Coulomb~interaction}^{2n-1}} > 0,
\end{equation}
which results in strong geometrical frustration.
They are similar to what has been found in
the quarter-filled extended Hubbard models on a two-leg ladder~\cite{vojta2001}
and a square lattice~\cite{mckenzie2001}.

 Analogously to spin exchange interactions, permutation terms in the present system
always appear through an even number of virtual processes.
As shown in Fig.~\ref{fig:UV_Jeff_Jr3},
three-site permutation terms appear
through six cyclic processes,
namely, $K_3^{\triangleright} = - d_3 t^6/V^5 + d'_3 t^6/(V^4U)$
[$K_3^{\triangleleft} = d''_3 t^6/(V^3U^2)$]
for triangles in (out of) hexagons
with $d_3$, $d'_3$, and $d''_3$ being positive constants.
Note that $K_3^{\triangleright} = - d_3 t^6/V^5$ survives even 
when $U=\infty$, which gives ferromagnetic interactions.
Moreover, four-site permutation terms appear through eight 
cyclic processes as shown in Fig.~\ref{fig:UV_Jeff_Jr4}, namely,
$K_4 = d_4 t^8/V^7 + d'_4 t^8/(V^6U) + d''_4 t^8/(V^5U^2) + d'''_4 t^8/(V^4U^3)$
with $d_4$, $d'_4$, $d''_4$, and $d'''_4$ being constants.

Since the three-spin permutation operator can be written as 
a product of two exchange operators
\begin{equation}
 P_3 = P_{ijk} = P_{ij} P_{ik} = 
\frac{1}{4}(1+4\bm{S}_i\cdot\bm{S}_j)(1+4\bm{S}_i\cdot\bm{S}_k)
\end{equation}
these terms become nearest-neighbor exchange interactions
on the effective triangular lattice~\cite{roger1983}
\begin{equation}
 P_3 + P_3^{-1} = 
\frac{1}{2}(1+4\bm{S}_i\cdot\bm{S}_j+4\bm{S}_j\cdot\bm{S}_k+4\bm{S}_k\cdot\bm{S}_i).
\end{equation}
$K_3^{\triangleright}$ and $K_3^{\triangleleft}$ will be 
renormalized into $J_1$, and the Hamiltonian
up to the constant term is rewritten as
\begin{eqnarray}
H &=& \sum_{\langle i,j\rangle_1} J_1^{\mathrm{eff}} \bm{S}_i\cdot\bm{S}_j 
 + \sum_{\langle i,j\rangle_2} J_2  \bm{S}_i\cdot\bm{S}_j 
 + \cdots
\nonumber
\\
 &&
 + \sum_{\square} K_{4} (P_{4}+P_{4}^{-1})
 + \cdots.
\label{eq_A_J_eff}
\end{eqnarray}
Here, $J_1^{\mathrm{eff}}$ is a linear combination of $J_1$,
$K_3^{\triangleright}$, and $K_3^{\triangleleft}$, namely,
$J_1^{\mathrm{eff}} = J_1 + 2K_3^{\triangleright} + 2K_3^{\triangleleft}$.

The effective nearest-neighbor interaction $J_1^{\mathrm{eff}}$ can be 
both ferromagnetic and antiferromagnetic depending on the size of 
$U/t$ and $V/t$, as shown in Fig.~\ref{fig:UV_Jeff_PD}.
When $U$ is moderately large and $V$ is extremely large [$U/t\ll (V/t)^3$],
$|J_1|\gg |K_3^{\triangleright}|, |K_3^{\triangleleft}|$ 
and hence $J_1^{\mathrm{eff}}\sim J_1 \propto t^4/(V^2U)$ is antiferromagnetic.
Since $|J_2|\sim |K_4|\sim t^8/(V^4U^3)$, the Hamiltonian effectively becomes an
antiferromagnetic $J_1^{\mathrm{eff}}$-$J_2$ Heisenberg model 
with four-spin ring exchange interaction $K_4$.
When $J_2$ is large enough, collinear antiferromagnetic order
overcomes 120$^\circ$ order~\cite{jolicoeur1990,chubukov1992}.

When $U$ is extremely large and $V$ is moderately large [$U/t\gg (V/t)^3$],
$|K_3^{\triangleright}|\gg |J_1|, |K_3^{\triangleleft}|$ and hence 
$J_1^{\mathrm{eff}}\sim K_3^{\triangleright} \propto -t^6/V^5$ is ferromagnetic.
Since $|J_1|\gg |K_4|\gg |J_2|$, the Hamiltonian effectively becomes a
ferromagnetic Heisenberg model with small four-spin ring-exchange $K_4$.

Finally, when $U/t\sim (V/t)^3$, $t^6/V^5$ terms in interactions $|J_1|$, 
$|K_3^{\triangleright}|$, and $|K_3^{\triangleleft}|$ nearly cancel out,
and $J_1^{\mathrm{eff}}$ becomes extremely small $\sim t^8/V^7$.
In this case, $J_2\sim t^{12}/V^{11}$ is much smaller than $J_1^{\mathrm{eff}}$,
while $K_4\sim t^8/V^7$ is comparable to $J_1^{\mathrm{eff}}$.
The Hamiltonian effectively becomes a four-spin ring-exchange model.

\section{Gutzwiller approximation results on the extended Hubbard model
on the honeycomb lattice at $n=3/2$}
\label{sec:results}

In this appendix we present  details on the calculations
of the phase diagram for the extended Hubbard model
on the honeycomb lattice at $n=3/2$ by employing the Gutzwiller
approximation.

\subsection{In the absence of nearest-neighbor Coulomb interaction ($V=0$)}
\label{sec:V=0}

\begin{figure}[t]
\includegraphics[width=0.9\columnwidth]{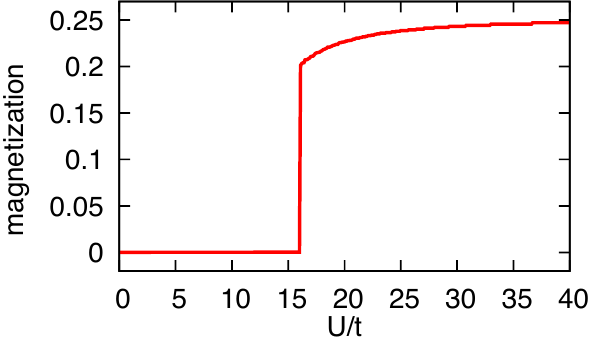}%
\caption{(Color online)
Magnetization for the metallic state at $V=0$
obtained by the Gutzwiller approximation for the
$n=3/2$ honeycomb lattice.
} 
\label{fig:V0_GA}
\end{figure}

We consider the case $U \neq 0$ and $V=0$ 
on the doped honeycomb model at $3/4$ filling 
where charge order and/or
ferromagnetism~\cite{nagaoka1966,hanisch1997}
are expected to be stable.

As we have discussed in the main text,
by using first the restricted Hartree-Fock method,
we find
the transition from a normal metal to a ferromagnetic
metal at $U/t\sim 5$.
When $U/t\gtrsim 6$, spins are fully polarized.
An up-spin band becomes fully occupied, while a down-spin band becomes
half occupied. Since the up- and down-spin bands are similar to the
original honeycomb band, the Fermi level is located at the Dirac node
of the down-spin band (semimetallic).

\begin{figure}[t]
\includegraphics[width=0.9\columnwidth]{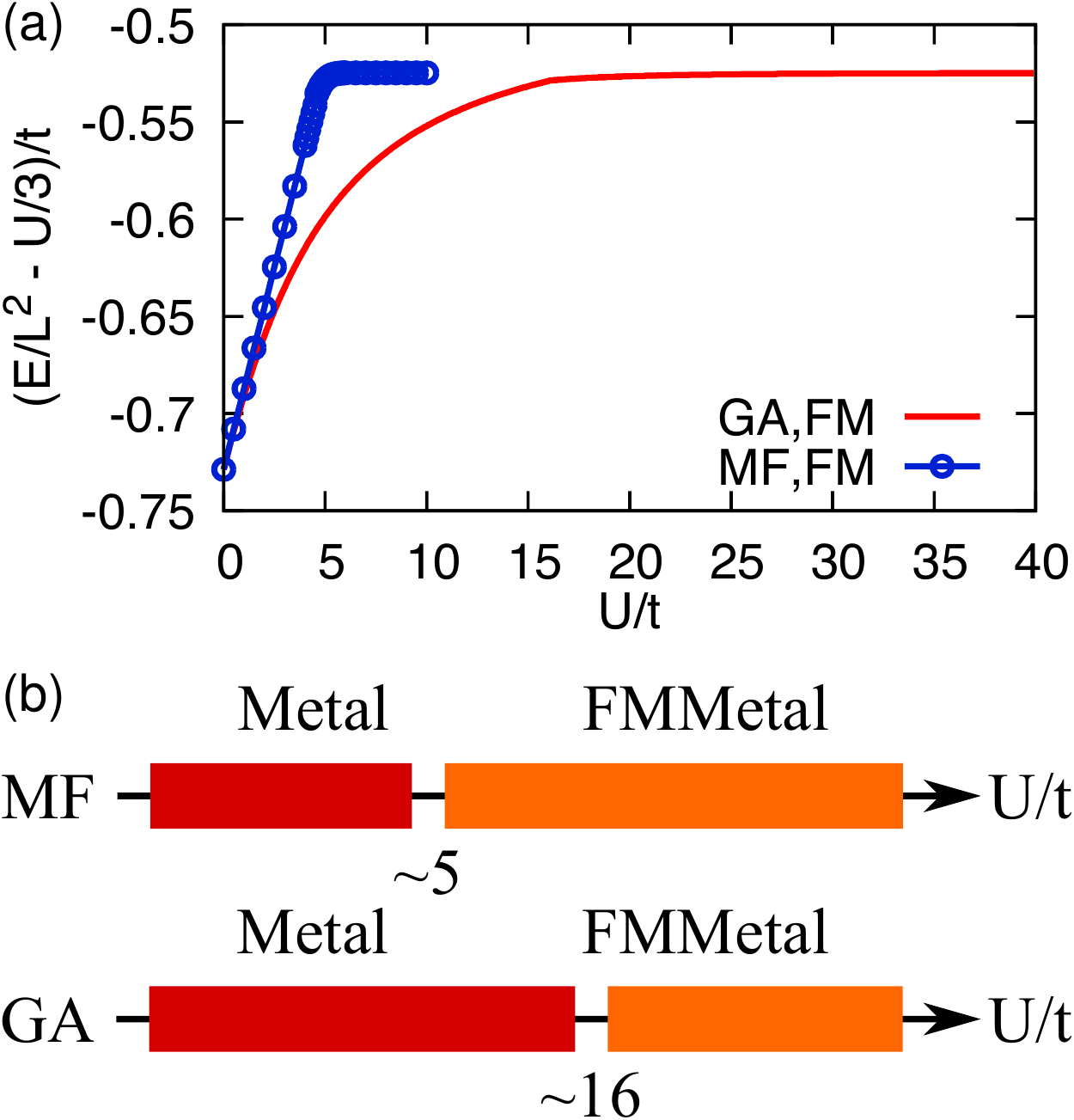}%
\caption{(Color online)
(a) Comparison of the energies at $V=0$ obtained for
the honeycomb lattice at $3/4$ filling by Hartree-Fock 
and by the Gutzwiller approximation. The energy is
saturated when the FM metal becomes fully polarized.
(b) Schematic phase diagrams obtained by each method.}
\label{fig:V0_PD}
\end{figure}

In order to assert the stability
of the ferromagnetic state  against 
quantum fluctuations beyond the mean-field treatment,
we apply the Gutzwiller approximation.
By assuming that the $c$ and $d$ orbitals are equivalent,
 we obtain the energy per bond as
\begin{widetext}
\begin{eqnarray}
  E(n_e,m,D)
 &=&
  \frac{U}{3}2D
  - t\left<T_{cd\uparrow}+T_{dc\uparrow}\right>_0
  \times
  \frac{(\sqrt{(1-2n_e+D)(n_e+m-D)} + \sqrt{(n_e-m-D)D})^2}
  {(n_e+m)(1-(n_e+m))}
\nonumber\\
 &&
  - t\left<T_{cd\downarrow}+T_{dc\downarrow}\right>_0
  \times
  \frac{(\sqrt{(1-2n_e+D)(n_e-m-D)} + \sqrt{(n_e+m-D)D})^2}
  {(n_e-m)(1-(n_e-m))},
\end{eqnarray}
\end{widetext}
where
$n_e = (n_{c,\uparrow}+n_{c,\downarrow})/2 =
(n_{d,\uparrow}+n_{d,\downarrow})/2 = 3/4$ is the number of electrons,
$m = (n_{c,\uparrow}-n_{c,\downarrow})/2 =
(n_{d,\uparrow}-n_{d,\downarrow})/2$ is the magnetization,
and $D = d_c = d_d$ is the number of doublons.
Here, $\left<T_{cd\sigma}+T_{dc\sigma}\right>_0$ is a function of
$n_{\sigma} = n_e + \sigma m$.
By numerically searching the energy minimum for $D\in[1/2,9/16]$ and $m\in[0,|n_e-D|]$,
we find a first-order transition from a normal metal to 
a ferromagnetic metal at $U/t\sim 16$,
as shown in Fig.~\ref{fig:V0_GA}. Inclusion of quantum fluctuations
as done in the Gutzwiller approximation
shifts the critical $U$ to larger values than in the 
Hartree-Fock approximation (see Fig.~\ref{fig:V0_PD}).
Quantum fluctuations seem to favor a metallic state without
ferromagnetism. 

\begin{figure}[t]
\includegraphics[width=0.9\columnwidth]{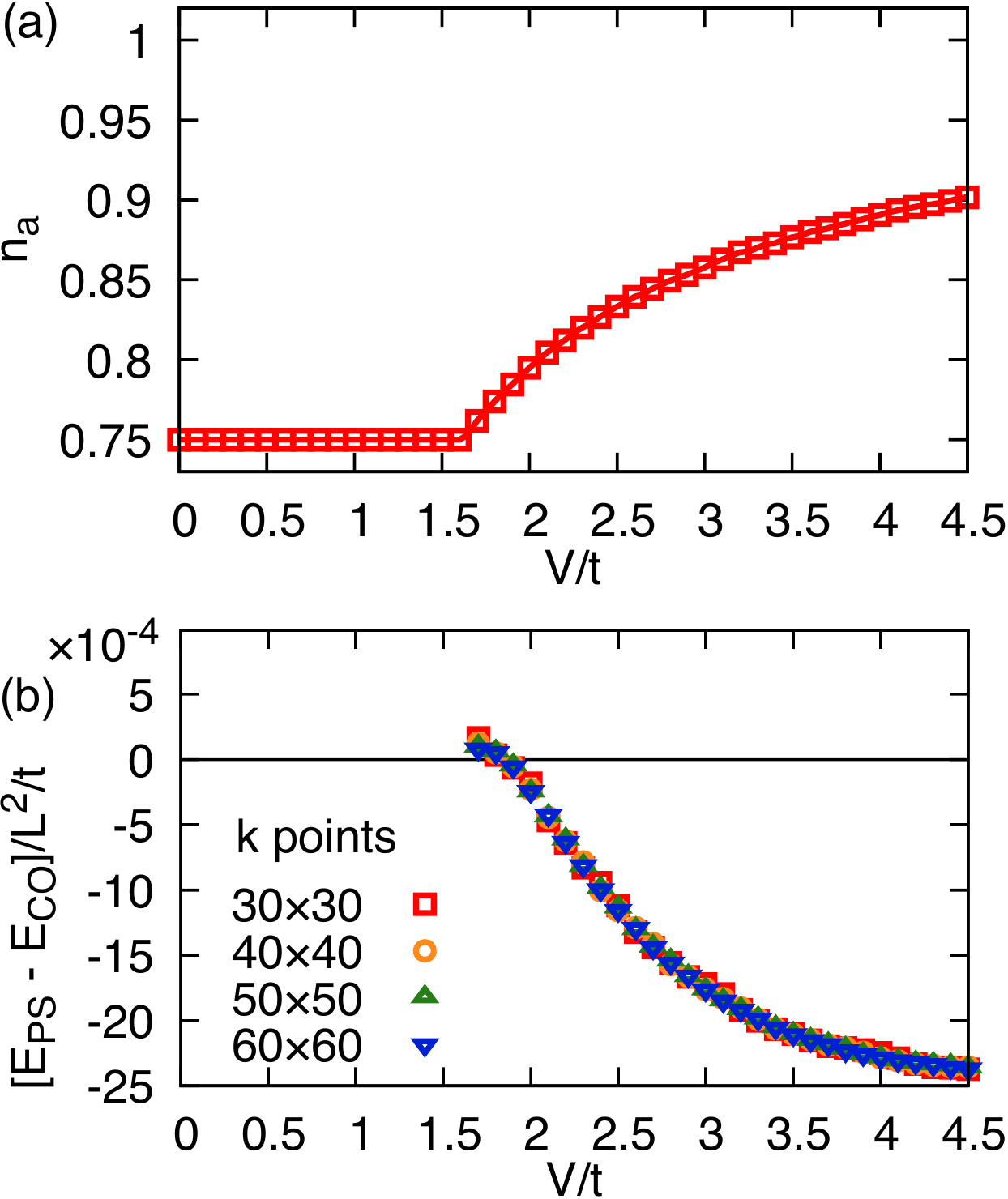}%
\caption{(Color online)
Results for the honeycomb lattice at $3/4$ filling
for $U=0$ obtained by the Gutzwiller
approximation.
(a) Charge order parameter $n_a$ as a function of $V/t$, from 
the uniform to the charge-ordered state. The number of electrons 
of charge-rich sites is given as $2n_a$.
(b) Energy difference between charge ordered and phase separated
states. They are nearly degenerate.}
\label{fig:U0_GA}
\end{figure}

\subsection{In the absence of on-site Coulomb interaction ($U=0$)}
\label{sec:U=0}

We focus here on possible charge ordered states and 
phase separation for $U=0$ and $V \neq 0$.

As we have discussed in the main text,
the restricted Hartree-Fock method finds
the transition from a normal metal to a
charge ordered metal at $V/t\sim 2$.
Charge-poor sites form an emergent triangular
structure as shown in Fig.~\ref{fig:MF_phase_diag} (c);
however, we find it to be nonmagnetic.

We now proceed with the Gutzwiller approximation.
Let us first focus on the uniform charge ordered state.
We assume the absence of magnetization:
\begin{eqnarray}
 n_{c\uparrow} &=& n_{c\downarrow} = n_{c\sigma},
\\
 n_{d\uparrow} &=& n_{d\downarrow} = n_{d\sigma}.
\end{eqnarray}
Since the total number of electrons is conserved ($n_{c\sigma}+n_{d\sigma}=3/4$),
the energy is given as a function of a single parameter $n_{c\sigma}$:
\begin{eqnarray}
 &&
  E(n_{c\sigma}) 
 =
  4Vn_{c\sigma}\left(\frac{3}{2}-n_{c\sigma}\right)
  - \frac{32}{3} t\left<T_{cd\sigma}+T_{dc\sigma}\right>_0
\nonumber\\
 &&
  \times
  \sqrt{n_{c\sigma}(1 - n_{c\sigma})
  \left(\frac{3}{2}-n_{c\sigma}\right)\left(n_{c\sigma}-\frac{1}{2}\right)}.
\end{eqnarray}
The expectation value of the hopping $\left<T_{cd\sigma}+T_{dc\sigma}\right>_0$ is a
constant since $n_{c\uparrow}+n_{d\uparrow} = n_{c\downarrow}+n_{d\downarrow} = 3/4$
is a constant. Hereafter, we abbreviate 
$t\left<T_{cd\sigma}+T_{dc\sigma}\right>_0|_{n_{c\sigma}+n_{d\sigma}=3/4}
=t_0\sim 0.36$.
By minimizing the energy for $n_{c\sigma}\in[3/4,1]$, we find a phase transition 
from a normal metal to a charge ordered metal at $V_{c}/t_0=40/9$.
The charge order parameter for $V>V_{c}$ is given as
\begin{equation}
 n_{c\sigma} = \frac{3}{4} + \frac{1}{2}
 \sqrt{\frac{5}{4} - \left[1-\left(\frac{8t_0}{3V}\right)^2\right]^{-1/2}}.
\end{equation}
In the metallic phase the energy is $E = (9/4)V - 2t_0$, while in the charge 
ordered phase the energy is $E = V\{1+\sqrt{1-[8t_0/(3V)]^2}\}$.
Thanks to the kinetic energy gain, the latter energy is lower than
that of the fully charge ordered (2121$\cdots$) insulating state ($E=2V$).

Next, we consider the possibility of phase separation consisting of
charge ordered insulator and metal phases.
We separate the system into two regions:
in the region $\kappa$ a fully charge ordered insulating state ($n_{c\sigma}=1$ and $n_{d\sigma}=0$) is realized,
while in the region ($1-\kappa$) a metallic state (the average number of electrons is $n_e$) is realized.
In the former region, there is no kinetic energy gain ($\langle T_{cd\sigma} \rangle = 0$)
and no intersite Coulomb energy loss ($E_V=0$ since $n_{d\sigma}=0$).
One  only has to consider energy in the latter region, which is given as
\begin{equation}
 E(\kappa,n_{e})
 =
  (1-\kappa) \left[ 4V n_{e}^2
   - 2t \left<T_{cd\sigma}+T_{dc\sigma}\right>_0
  \right].
\end{equation}
The conservation of charge yields
\begin{eqnarray}
 2\kappa + (2n_{e}+2n_{e})(1-\kappa) &=& 3,
\\
 \kappa = 1 - \frac{1}{4n_{e}-2},
\end{eqnarray}
which simplifies the energy
\begin{equation}
 E(n_{e})
 =
  \frac{1}{2n_{e}-1} \left[ 2V n_{e}^2
   - t \left<T_{cd\sigma}+T_{dc\sigma}\right>_0
  \right].
\end{equation}
Here, $\left<T_{cd\sigma}+T_{dc\sigma}\right>_0$ is a function of $n_e$.
If the energy is minimized for $\kappa>0$, phase separation takes place.
By numerically searching the energy minimum for $n_e\in[3/4,1]$ ($\kappa\in[0,1/2]$),
we find a phase separated state for $V>V_c$. The charge ordered and phase 
separated states are found to be nearly degenerate, as shown in 
Fig.~\ref{fig:U0_GA}, with the energy of the phase separated state 
being slightly lower.


\begin{thebibliography}{99}
\bibitem{imada1998}
M. Imada, A. Fujimori, and Y. Tokura, Rev. Mod. Phys. {\bf 70}, 1039 (1998).
\bibitem{baskaran2016}
G. Baskaran, Supercond. Sci. Technol. {\bf 29}, 124002 (2016).
\bibitem{watanabe2005}
H. Watanabe and M. Ogata, J. Phys. Soc. Jpn. {\bf 74}, 2901 (2005).
\bibitem{tocchio2014}
L.F. Tocchio, C. Gros, X.-F. Zhang, and S. Eggert, Phys. Rev. Lett. {\bf 113}, 246405 (2014).
\bibitem{kaneko2016}
R. Kaneko, L.F. Tocchio, R. Valent\'i, F. Becca, and C. Gros,
Phys. Rev. B {\bf 93}, 125127 (2016).
\bibitem{reja2015}
S. Reja, R. Ray, J. van den Brink, and S. Kumar, Phys. Rev. B {\bf 91}, 140403(R) (2015).
\bibitem{hotta2006} C. Hotta and N. Furukawa, Phys. Rev. B {\bf 74}, 193107 (2006).
\bibitem{miyazaki2009}
M. Miyazaki, C. Hotta, S. Miyahara, K. Matsuda, and N. Furukawa, J. Phys. Soc. Jpn. {\bf 78}, 014707 (2009).
\bibitem{canocortes2011}
L. Cano-Cortes, A. Ralko, C. Fevrier, J. Merino, and S. Fratini, Phys. Rev. B {\bf 84}, 155115 (2011).
\bibitem{merino2013}
J. Merino, A. Ralko, and S. Fratini, Phys. Rev. Lett. {\bf 111}, 126403 (2013).
\bibitem{akagi2015}
Y. Akagi and Y. Motome, Phys. Rev. B {\bf 91}, 155132 (2015).
\bibitem{wen2010}
J. Wen, A. Ruegg, C.-C. J. Wang, and G. A. Fiete, Phys. Rev. B {\bf 82}, 075125 (2010).
\bibitem{ferhat2014}
K. Ferhat and A. Ralko, Phys. Rev. B {\bf 89}, 155141 (2014).
\bibitem{pollmann2014}
F. Pollmann, K. Roychowdhury, C. Hotta, and K. Penc, Phys. Rev. B {\bf 90}, 035118 (2014).
\bibitem{roychowdhury2015}
K. Roychowdhury, S. Bhattacharjee, and F. Pollmann, Phys. Rev. B {\bf 92}, 075141 (2015).
\bibitem{wawrzynska2007}
E. Wawrzy\'{n}ska, R. Coldea, E. M. Wheeler, I. I. Mazin, M. D. Johannes, T.
S\"{o}rgel, M. Jansen, R. M. Ibberson, and P. G. Radaelli,
Phys. Rev. Lett. {\bf 99}, 157204 (2007).
\bibitem{wawrzynska2008}
E. Wawrzy\'{n}ska, R. Coldea, E. M. Wheeler, T. S\"{o}rgel, M. Jansen,
R. M. Ibberson, P. G. Radaelli, and M. M. Koza,
Phys. Rev. B {\bf 77}, 094439 (2008).
\bibitem{fulde2001}
P. Fulde, A. N. Yaresko, A. A. Zvyagin, and Y. Grin,
Europhys. Lett. {\bf 54}, 779 (2001).
\bibitem{fulde2002}
P. Fulde, K. Penc, and N. Shannon,
Ann. Phys. {\bf 11}, 892 (2002).
\bibitem{hayami2014}
S. Hayami, T. Misawa, Y. Motome, JPS Conf. Proc. {\bf 3}, 016016 (2014).
\bibitem{meng2010}
Z. Y. Meng, T. C. Lang, S. Wessel, F. F. Assaad, and A. Muramatsu,
Nature {\bf 464}, 847 (2010).
\bibitem{sorella2012}
S. Sorella, Y. Otsuka, and S. Yunoki,
Sci. Rep. {\bf 2}, 992 (2012).
\bibitem{raghu2008}
S. Raghu, X.-L. Qi, C. Honerkamp, and S.-C. Zhang, Phys. Rev. Lett. {\bf 100}, 156401 (2008).
\bibitem{weeks2010}
C. Weeks and M. Franz, Phys. Rev. B {\bf 81}, 085105 (2010).
\bibitem{capponi2015}
S. Capponi and A. M. L\"{a}uchli, Phys. Rev. B {\bf 92}, 085146 (2015).
\bibitem{motruk2015}
J. Motruk, A.G. Grushin, F. de Juan, and F. Pollmann, Phys. Rev. B {\bf 92}, 085147 (2015).
\bibitem{scherer2015}
D.D. Scherer, M.M. Scherer, and C. Honerkamp, Phys. Rev. B {\bf 92}, 155137 (2015).
\bibitem{kurita2015}
M. Kurita, Y. Yamaji, M. Imada, Phys. Rev. B {\bf 94}, 125131 (2016).
\bibitem{castro2011}
E.V. Castro, A.G. Grushin, B. Valenzuela, M.A.H. Vozmediano, A. Cortijo, and F. de Juan, Phys. Rev. Lett. {\bf 107}, 106402 (2011).
\bibitem{grushin2013}
A.G. Grushin, E.V. Castro, A. Cortijo, F. de Juan, M.A.H. Vozmediano, and B. Valenzuela, Phys. Rev. B {\bf 87}, 085136 (2013).
\bibitem{hanisch1997}
T. Hanisch, G.S. Uhrig, and E. M\"{u}ller-Hartmann, Phys. Rev. B {\bf 56}, 13960 (1997).
\bibitem{pasrija2016}
K. Pasrija and S. Kumar, Phys. Rev. B {\bf 93}, 195110 (2016).
\bibitem{jolicoeur1990} 
Th. Jolicoeur, E. Dagotto, E. Gagliano, and S. Bacci, Phys. Rev. B 42, 4800(R) (1990).
\bibitem{chubukov1992}
A. V. Chubukov and Th. Jolicoeur, Phys. Rev. B {\bf 46}, 11137 (1992).
\bibitem{roger1983}
M. Roger, J.H. Hetherington, and J.M. Delrieu, Rev. Mod. Phys. 55, {\bf 1} (1983).
\bibitem{motrunich2005}
O.I. Motrunich, Phys. Rev. B {\bf 72}, 045105 (2005).
\bibitem{grover2010}
T. Grover, N. Trivedi, T. Senthil, and P. A. Lee, Phys. Rev. B {\bf 81}, 245121 (2010).
\bibitem{holt2014}
M. Holt, B.J. Powell, and J. Merino, Phys. Rev. B {\bf 89}, 174415 (2014)
\bibitem{korshunov1993}
S.E. Korshunov, Phys. Rev. B {\bf 47}, 6165(R) (1993).
\bibitem{kubo1997}
K. Kubo, T. Momoi, Z. Phys. {\bf B} 103, 485 (1997).
\bibitem{becca2000}
F. Becca, M. Capone, and S. Sorella, Phys. Rev. B {\bf 62}, 12700 (2000).
\bibitem{anderson1988}
P.W. Anderson, G. Baskaran, Z. Zou, J. Wheatley, T. Hsu, B.S. Shastry, B. Doucot, S. Liang,
Physica C {\bf 153}, 527 (1988).
\bibitem{balents2010}
L. Balents, Nature {\bf 464}, 199 (2010).
\bibitem{poilblanc2009}
D. Poilblanc and H. Tsunetsugu,
\emph{Mobile Holes in Frustrated Quantum Magnets and Itinerant Fermions
on Frustrated Geometries},
Springer Series in Solid-State Sciences,
edited by Claudine Lacroix, Philippe Mendels, and Frederic Mila,
Vol. 164 (Springer, Berlin, Heidelberg, 2011), p. 563.
\bibitem{mila1993}
F. Mila and X. Zotos, Europhys. Lett. {\bf 24}, 133 (1993).
\bibitem{penc1994}
K. Penc and F. Mila, \prb {\bf 49}, 9670 (1994).
\bibitem{vojta2001}
M. Vojta, A. H\"{u}bsch, and R.M. Noack, Phys. Rev. B {\bf 63}, 045105 (2001).
\bibitem{tocchio2011}
L.F. Tocchio, F. Becca, and C. Gros, Phys. Rev. B {\bf 83}, 195138 (2011).
\bibitem{iqbal2013}
Y. Iqbal, F. Becca, S. Sorella, and D. Poilblanc, Phys. Rev. B {\bf 87}, 060405(R) (2013).
\bibitem{depenbrock2012}
S. Depenbrock, I.P. McCulloch, and U. Schollw\"{o}ck, Phys. Rev. Lett. {\bf 109}, 067201 (2012).
\bibitem{nagaoka1966}
Y. Nagaoka, Phys. Rev. {\bf 147}, 392 (1966).
\bibitem{sugita2016}
Y. Sugita and Y. Motome, J. Phys. Soc. Jpn. {\bf 85}, 073709 (2016).
\bibitem{tahara2008}
D. Tahara and M. Imada, J. Phys. Soc. Jpn. {\bf 77}, 114701 (2008).
\bibitem{ogawa1975}
T. Ogawa, K. Kanda, and T. Matsubara, Prog. Theo. Phys. {\bf 53}, 614 (1975).
\bibitem{zhang1988}
F.C. Zhang, C. Gros, T. M. Rice, and H. Shiba, Supercond. Sci. Thechnol. {\bf 1} 36 (1988).
\bibitem{gros1988}
C. Gros, Phys. Rev. B {\bf 38}, 931(R) (1988).
\bibitem{capello2005}
M. Capello, F. Becca, M. Fabrizio, S. Sorella, and E. Tosatti, Phys. Rev. Lett. {\bf 94}, 026406 (2005).
\bibitem{tocchio2008}
L.F. Tocchio, F. Becca, A. Parola, and S. Sorella, Phys. Rev. B {\bf 78}, 041101(R) (2008).
\bibitem{yunoki2006}
S. Yunoki and S. Sorella, Phys. Rev. B {\bf 74}, 014408 (2006).
\bibitem{mckenzie2001}
R.H. McKenzie, J. Merino, J.B. Marston, and O.P. Sushkov, Phys. Rev. B {\bf 64}, 085109 (2001).
\end{thebibliography}
\end{document}